\documentclass{aa}
\usepackage{txfonts}
\usepackage{graphicx}
\usepackage[usenames]{color}

\newcommand{\teff}{$T_{\rm eff}$}
\newcommand{\logg}{log\,$g$}
\newcommand{\vsini}{$v$\,sin\,$i$}

\begin{document}

\title{A high resolution study of isotopic composition and 
chemical abundances of blue horizontal branch stars in the globular clusters NGC\,6397 and
NGC\,6752\thanks{Based on observations obtained at the European Southern Observatory,
Paranal, Chile (ESO programmes 076.D-0169(A) and 081.D-0498(A)).}}
\titlerunning{Isotopes in BHB stars}

\author{
S.~Hubrig
\inst{1,2}
\and
F.~Castelli
\inst{3}
\and
G.~De~Silva
\inst{4}
\and
J.~F.~Gonz\'alez
\inst{5}
\and
Y.~Momany
\inst{1}
\and
M.~Netopil
\inst{6}
\and
S. Moehler\inst{4}
}
%\offprints{F. Castelli}

\institute{
Astrophysical Institute Potsdam, An der Sternwarte 16, D-14482 Potsdam, Germany\\
\email{shubrig@aip.de}
\and
European Southern Observatory, Casilla 19001, Santiago 19, Chile
\and
Istituto Nazionale di Astrofisica--
Osservatorio Astronomico di Trieste, Via Tiepolo 11,
I-34131 Trieste, Italy
%\email{castelli@oats.inaf.it}
\and
European Southern Observatory, Karl-Schwarzschild-Str.\ 2, 85748 Garching, Germany
\and   
Complejo Astron\'omico El Leoncito, Casilla 467, 5400 San Juan, Argentina
\and
Institut f\"ur Astronomie der Universit\"at Wien, T\"urkenschanzstr. 17, 1180 Wien, Austria
}

\date{}

\abstract
{}
{
Large abundance anomalies have been previously detected in Horizontal Branch B-type stars.
It has been suggested that at \teff{}$>$11\,000\,K the stellar atmospheres become 
susceptible to diffusion effects and thus develop surface abundances similar to those 
that appear in main-sequence chemically peculiar group of A and B-type stars.
We present the first high resolution study of isotopic anomalies and chemical abundances 
in six Horizontal Branch B-type stars in globular clusters NGC\,6397 and NGC\,6752 
and compare them to those observed in HgMn stars.
% and the Pop~II halo B-type star Feige\,86.
}
{
We obtained high-resolution (up to R$\sim$115\,000) UVES spectra of a 
representative sample of six B-type stars (T183, T191, T193, B652, B2151, B2206) with sharp 
spectral lines (\vsini{}$\le$10\,km\,s$^{-1}$).
The stars, T183 and B2151, were observed on two consecutive nights to examine potential spectrum variability.
A detailed spectrum analysis is done relying on Kurucz 
ATLAS9 and ATLAS12 models. The spectra are analysed using the SYNTHE code to generate synthetic spectra.
}
{
It is the first time an abundance analysis is performed 
for all elements for which spectral lines were detected in UVES spectra of Horizontal Branch B-type stars.
Our study of these stars revealed no signs of He isotopic anomalies which would produce a
$^{3}$He/$^{4}$He ratio different from the solar one.
The isotopic anomaly of Ca is detected in all six studied stars.
The chemical abundance analysis reveals an overabundance over the solar values 
of P, Ti, Mn, Fe, and Y and an
overabundance over the cluster metallicity of Mg, Ca, and Cr.
This behaviour is very similar in all 
six studied stars of both clusters with a few exceptions: The Na abundance 
%in B652 and B2206 
is by more than 1.4\,dex larger than the cluster metallicity in B652, and by more than 0.8\,dex larger than the 
cluster metallicity in B2206;
the Co abundance is 1.0\,dex over the solar abundance for T191, while Zr is overabundant  over the solar 
abundance by 0.4\,dex in B2206.
No lines of Hg or other heavy elements were observed in the spectra. Weak emission lines of 
\ion{Ti}{ii}, similar to those frequently observed in HgMn stars were discovered in one Horizontal 
Branch B-type star T191. Further, we detected a radial velocity change of 0.9\,km\,s$^{-1}$ from one night to the next
for T183 and 0.4\,km\,s$^{-1}$ for B2151. Further high-resolution observations are necessary to examine whether both 
stars are spectroscopic binaries.}
{}


\maketitle{}

\section{Introduction}

It is generally thought that isotopic anomalies are restricted 
to the chemically peculiar (CP) group of A and B-type stars.
The study of isotopic anomalies of different chemical elements in chemically peculiar stars dates back to 
the 1960s when Sargent \& Jugaku (\cite{SargentJugaku1961}) detected an anomalously high concentration of 
$^3$He in the chemically peculiar Bp star 3\,Cen\,A.
Chemically peculiar A and B stars are main-sequence stars in the spectra of which lines of a 
number of elements appear abnormally strong or weak with respect to the bulk of normal A and B dwarfs
of the same temperature. Altogether, the CP stars include the He-rich, He-weak, HgMn, Si, SrCrEu, and 
Am stars. The remarkable variety of elemental overabundances and depletions in the atmospheres of CP stars 
has been a puzzle for stellar spectroscopists for more than a century. Ap and Bp stars with significant 
dipolar or quadrupolar stellar magnetic fields have overabundances
of Si, Ti, Cr, Sr, and rare earths, of up to several dex with underabundances of other elements such as He and O. Am stars 
show mild overabundances of most iron peak elements, but always abnormally weak \ion{Ca}{ii} lines. The
most puzzling CP stars with amazing chemical and isotopic anomalies, so called HgMn stars, 
belong to a group of most slowly rotating late B-type stars and show substantial overabundances of 
a number of elements, among them heavy ones, such as Pt, Au, Hg, Tl, and Bi,
and isotopic anomalies with pattern changing from one star to the next (Hubrig et al.\ \cite{Hubrig1999}).
In the magnetic Ap and Bp stars the rare earth elements are among the most 
enhanced, up to 3--4\,dex, whereas the heavy elements are overabundant in the atmospheres of HgMn 
stars by up to a million times compared to the solar system level.

The mechanisms responsible for producing the chemical peculiarities in upper main-sequence
stars are still unclear. It is generally believed that the peculiar atmospheric abundances are a 
consequence of trace element diffusion from the stellar envelope.
The anomalous abundances probably arise from competition between gravitational settling, radiative 
levitation, turbulent and convective mixing, and mixed and fractionated mass loss (e.g., Gonzales et al.\ \cite{Gonzales1995}).
However, the observed
 abundance distributions in these stars have never been completely explained by any calculation. 
The radiative diffusion  mechanism alone
can not account for all abundance and isotopic anomalies (cf.\ Proffitt et al.\ \cite{Proffitt1999}). 
Optical studies based on high resolution spectra of HgMn stars concentrated
on the behaviour of the \ion{Hg}{ii} $\lambda\,3984$, \ion{Hg}{i} $\lambda\,4385$, and  \ion{Hg}{ii} $\lambda\,6149$ 
lines (e.g.\ Dolk et al.\ \cite{Dolk2003}). 
The high resolution spectroscopic study of the isotopic composition of  Pt  (Hubrig et al.\ \cite{Hubrig1999})
 showed that this element behaves very similar to Hg, i.e.\ cooler 
stars exhibit a large relative overabundance of the heavy isotopes. 
%However, for some stars a very unusual Hg isotopic structure has been reported. 

A new isotopic abundance anomaly in HgMn and Ap stars was discovered a few 
years ago by Castelli \& Hubrig (\cite{CastelliHubrig2004a}) and Cowley \& Hubrig (\cite{CowleyHubrig2005}).
The observational evidence for large isotopic shifts in the infrared triplet of  
\ion{Ca}{ii} has been presented in a sample of HgMn stars (Castelli \& Hubrig \cite{CastelliHubrig2004a}).
Calcium has six stable isotopes with mass numbers 40, 42, 43, 44, 46, and 48.
The terrestrial isotopic mixture (in percent) is 96.941, 0.647, 0.135, 2.086, 0.004, and 
0.187, respectively (Anders \& Grevesse \cite{AndersGrevesse1989}).
The maximum isotopic shifts between the isotopes $^{40}$Ca and $^{48}$Ca 
for the \ion{Ca}{ii} lines at 3968.47\,\AA{} (H-line) and at 3933.64\,\AA{} (K-line) is
of the order of only $\sim$9\,m\AA{} (Martensson-Pendril et al.\ \cite{Martensson-Pendril1992}), 
so that these lines are practically unaffected by possible calcium isotopic composition anomalies.
On the other hand, the  largest isotopic shift in the \ion{Ca}{ii}  infrared triplet at 
8498.023\,\AA{}, 8542.091\,\AA{}, and 8662.141\,\AA{}, 
due to $3{\rm d}^2{\rm D}\,\rightarrow\,4{\rm p}^2{\rm P}$ transitions,
is about 0.2\,\AA{}, according to N\"ortersh\"auser et al.\ (\cite{Noertershaeuser1998}).
%Among the studied stars, the record holder is the HgMn star HD\,175640 for which 
%the measured wavelength is consistent with \ion{Ca}{ii} being present in the atmosphere entirely 
%in form of the heaviest stable isotope $^{48}$Ca (Fig.~\ref{fig:si_fit}). This is a very striking result 
%as $^{48}$Ca makes up only 0.187\% of the terrestrial Ca mixture.

As the origin of these isotopic anomalies is not known with certainty, it is important to investigate 
the isotopic behaviour in a variety of stars. 
%The discovery of this phenomenon in CP stars does not necessarily mean that it is restricted to this group. 
The best candidates to search for similar chemical anomalies are slowly rotating blue Horizontal 
Branch B-type (BHB) stars, which have  
similar \teff{} and \logg{} as HgMn stars and shallow surface convective zones.
The so-called hot BHB stars with \teff{}$>$11\,000\,K display surface abundances which are 
significantly altered by diffusion effects in their atmospheres (e.g.\ Behr \cite{Behr2003}).
Stars in the horizontal branch (HB) have already passed the red giant branch (RGB) phase and are 
presently burning helium in their core (Hoyle \& Schwarzschild \cite{hoyle1955}).
In a number of globular clusters the HB show ``jumps'' where stars appear brighter than theoretical
expectations (e.g., Grundahl et al.\ \cite{Grundahl1999};
Momany et al.\ \cite{Momany2002}, \cite{Momany2004}), or 
gaps, i.e.\ underpopulated regions along the HB (e.g.\ Sosin et al.\ \cite{sosin1997}).
While in cooler HB stars with \teff{}$<$11\,200\,K metal abundances are in rough agreements 
with the canonical cluster metallicity, most of stars with \teff{}$\gtrsim$11\,200\,K are helium depleted and 
show an increase of their metal content (e.g.\ Behr et al.\
\cite{Behr2000}, \cite{Behr2003}).
Since almost all of these hot BHB stars are slowly rotating, it is expected that the observed abundance anomalies
are developed due to diffusion effects. 

Similarly, a few mid-B spectral type stars are known to lie close to or on the horizontal branch, 
such as Feige\,86, PHL\,25, PHL\,1434, and HD\,135485. Abundance anomalies in these 
field BHB stars have been detected in various studies (e.g., Trundle et al.\ \cite{trundle2001}; 
Bonifacio et al.\ \cite{Bonifacio95}), which are most probably 
caused by radiative levitation and gravitational settling.
Since April 2006 we have at our disposal a 
high resolution spectrum of the Pop~II halo B-type star Feige\,86 for which, for the first time, we have 
been able to determine the isotopic structure of the \ion{Hg}{ii} $\lambda\,3984$ line and a \ion{Ca}{ii} triplet shift 
of 0.13\,\AA{}. The \ion{He}{i} lines have been previously studied by 
Bonifacio et al.\ (\cite{Bonifacio95}), and the detection of $^3$He at 6678\,\AA{} was announced twenty years ago
by Hartoog (\cite{Hartoog1979}).
Similar to Feige\,86, and to many HgMn stars, BHB stars show overabundant 
P, Mn, Fe Ti, and Cr and underabundant He (e.g., Fabbian et al.\ \cite{Fabbian2005}). However, the 
{\it isotopic} composition of He, Ca, and heavy elements in this type of stars has never been 
studied before. 

In this work we focus our attention on the high resolution study of the isotopic composition and chemical 
abundances of various elements in BHB stars in the globular clusters NGC\,6397 and  NGC\,6752,
and compare them to those determined previously in HgMn stars and the Pop~II halo B-type 
star Feige\,86. All previous spectroscopic studies of BHB stars made use of much lower spectral 
resolution observation with the highest spectral resolution achieved with the HIRES spectrograph 
at the Keck 1 telescope (Behr \cite{Behr2003}).
%The position of the selected stars in 
%the CMD is presented in Fig.~\ref{fig:emis_mn}. 
The clusters  NGC\,6397 and  NGC\,6752 are the nearest globular clusters for which the phenomenon of
radiative levitation of iron and heavy elements has been detected with certainty and for which a
significant number of targets exist. 
% -- this makes it the best suited target cluster. 

\section{Sample selection and observations}

Understandably, a study of chemical abundances and especially of the isotopic composition in 
slowly rotating BHB stars 
requires the availability of high-resolution spectra observed at a 
reasonable signal-to-noise ratio (S/N).
Moderately high resolution spectra with a resolving power 
of 30\,000 to 40\,000 of a sample of BHB stars in NGC\,6397 and  NGC\,6752 were previously 
obtained at ESO in service mode with the VLT UV-Visual Echelle 
Spectrograph UVES at UT2 in the framework of programs 65.L-0233(A) and 69.D-0220(A). 
%With an agreement of the PI of these programs, S. Moehler, 
We retrieved these spectra from the ESO archive, to select the 
best suitable candidates with the most sharp-lined spectra and strong appearance of iron group elements. 
Based on this study, we selected  three BHB targets in NGC\,6397 (T183, T191, and T193)
and three BHB targets in NGC\,6752 (B665, B2151, and B2206).

\begin{figure}
\begin{center}
\includegraphics[width=0.55\textwidth,angle=0,clip=]{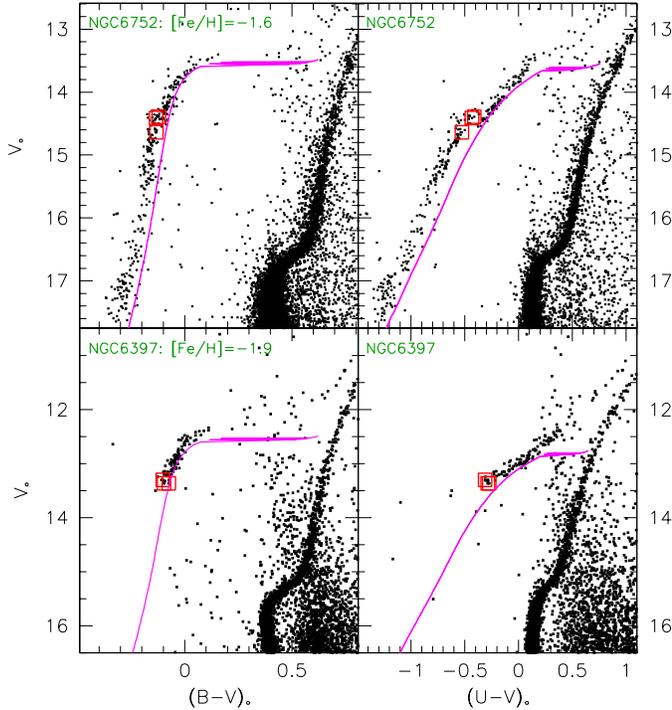}
\end{center}
\caption{Left   panels  display   the  classical   $V$,($B-V$)
color-magnitude diagrams  of NGC\,6397  (lower left) and  NGC\,6752 (upper
left). For both  clusters, the studied BHB samples  are plotted as open
squares. Appropriate metallicity ZAHB models from Pietrinferni et al.\
\cite{Pietrinferni2006}) are  also  plotted.    Right  panels  display the respective
ultraviolet $V$,($U-V$) diagrams.  In  the case of NGC\,6752, the location
of the Grundahl  et al.\ (\cite{Grundahl1999}) $u$-jump is easily distinguishable at
($U-V$)$\simeq-0.4$.
}
\label{fig:col_mag}
\end{figure}

Further we checked the position of these targets in the color-magnitude diagrams of NGC6397 and NGC6752.
In   Fig.~\ref{fig:col_mag}  we   show  the dereddened  $V$,($U-V$)   and  $V$,($B-V$)
color-magnitude diagrams of NGC6397 and NGC\,6752 (Momany et  al.\  \cite{Momany2003}),
and  highlight the  location of the selected BHB stars by open  squares.
The photometric data of NGC\,6397 and NGC\,6752 were
obtained from ground-based
$UBV$  observations collected in three different  runs with the  Wide-Field
Imager  (WFI) at  the 2.2\,m  ESO-MPI telescope, and were
previously published in Momany et al.\ (\cite{Momany2002}; \cite{Momany2003}).
Contrary  to the classical  $V$,($B-V$) diagrams (left
panels) where  the color saturation  results in an almost  vertical HB
sequence,  the ultraviolet  diagrams (right  panels)  show interesting
photometric features as stars become hotter along the HB.

Grundahl  et al.\ (\cite{Grundahl1999}) identified a  discontinuity in  the Stromgren
($u$,$u-y$) diagram at around  11\,500~K.  This discontinuity is easily
detectable in the upper right panel of Fig.~\ref{fig:col_mag}
[($U-V$)$\simeq-0.4$], and  marks a sudden divergence  with respect to
Zero-Age-Horizontal-Branch   (ZAHB)   models, i.e. stars   between
11\,500--20\,000~K  appear  brighter. This discontinuity was
attributed to  radiative levitation of heavy  elements that eventually
enhance the atmospheric abundances  of these BHB stars. Noteworthy,
this $u-$discontinuity  was proved to be present in all BHB clusters,
occurring  always at a similar temperature regardless of the cluster
parameters (e.g., age, metallicity, etc.). In  particular, Grundahl et
al.  find  correspondence between the discontinuity  and the so-called
gravity-jump  (observed in  both globular clusters and field stars).
Furthermore,  Recio-Blanco et  al.\ (\cite{RecioBlanco2002})  show  that the
$u$-discontinuity might be related to the discontinuity in rotational velocity
found around \teff{}$\ge$10\,000\,K (see also Behr
et al.\ \cite{Behr2000}).
Thus, the triggering  of the  Grundahl  et al.\  $u$-jump traces  the
imprint of other anomalies (heavy element abundances,  gravity, and rotational
velocity).  In this  context, it is expected 
that the selected BHB stars  are post-$u$ jump,
and  therefore exhibit the above mentioned anomalies.

In the case of the NGC\,6752  BHB star sample this is a straightforward
conclusion. Indeed, in the upper  right panel of  Fig.~\ref{fig:col_mag} we
compare the  BHB star  distribution to that  of a  [Fe/H] = $-1.6$ ZAHB
model (Pietrinferni  et al.\ \cite{Pietrinferni2006}), and show  that all three
selected stars are brighter
than the ZAHB model. On the  other hand, the horizontal branch of NGC\,6397 is clearly
less  extended than  that of NGC\,6752, and  even  when the  $V$,($U-V$)
diagram  is used,  it is not easy to establish whether the 
selected stars are post-$u$ jump stars or not.

\begin{figure}
\begin{center}
\includegraphics[width=0.55\textwidth,angle=0,clip=]{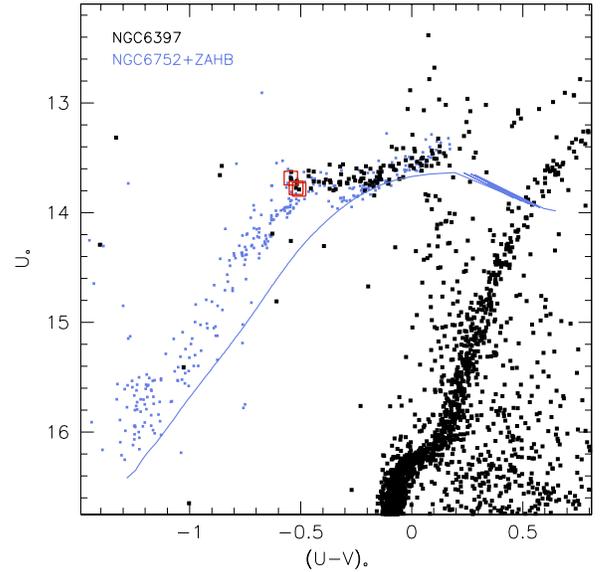}
\end{center}
\caption{
A superposition of the NGC6397 $U$,($U-V$) color-magnitude
diagram  on  the HB location  of NGC\,6752.   The ZAHB model has the
metallicity of NGC\,6752.  The superposition shows that three selected NGC\,6397 BHB stars
are most likely located on the post-$u$ jump HB segment.}
\label{fig:col_2mag}
\end{figure}

In Fig.~\ref{fig:col_2mag} we overplot the $U$,($U-V$)
diagram of the entire  NGC\,6752 HB sample  (grey squares) and
compare it with that of  NGC\,6397 (black  squares). The NGC\,6397
photometry has been shifted in such a way as to match the location of the NGC\,6752
HB segment between  the Grundahl et al.\ jump  and the blue-boundary of
the instability strip.  Keeping in mind that the
$u$-jump has been found to be an ubiquitous feature occurring always at
a similar  temperature for all globular clusters (regardless of their
metallicity and other properties) we conclude that such a superposition  
technique  is helpful  in  highlighting  the segmented nature of the globular cluster  
BHB stars (see e.g. Figure~7 of Momany et  al. \cite{Momany2004}).
Fig.~\ref{fig:col_2mag} shows that three selected NGC\,6397 BHB stars 
are most likely located on the post-$u$ jump HB segment.

\begin{table*} 
\caption{$UBV$ magnitudes of  the NGC\,6752  and NGC\,6397
BHB star sample.  The first  identification numbers (Star ID), positions, and
$UBV$  photometry are from Momany  et al.\ \cite{Momany2003}, while  the second
identification  numbers are from  Buonanno et  al.\ (\cite{Buonanno1986}) and 
Anthony-Twarog \& Twarog\ (\cite{AntonyTwarog2000}) for NGC\,6397  NGC\,6752, respectively.
Radial velocities $v_{\rm rad}$ and values of \vsini{} are listed in the last two columns.}
%is obtained from the superposition of observed 
%spectra with the synthetic ones. It stands for the correction in the radial velocity and heliocentric velocity}
\label{tab:photometry}
\centering
\begin{tabular}{rlrrlcrrrr}
\hline
\hline
\noalign{\smallskip}
\multicolumn{1}{c}{Cluster}&
\multicolumn{1}{c}{Star ID}&
\multicolumn{1}{c}{Star ID}&
\multicolumn{1}{c}{RA  (J2000)}&
\multicolumn{1}{c}{Dec (J2000)}&
\multicolumn{1}{c}{U}&
\multicolumn{1}{c}{B}&
\multicolumn{1}{c}{V}&
\multicolumn{1}{c}{$v_{\rm rad}$} &
\multicolumn{1}{c}{\vsini{}} \\
 & & & & & & & &
\multicolumn{1}{c}{[km\,s$^{-1}$]} &
\multicolumn{1}{c}{[km\,s$^{-1}$]} \\
%\multicolumn{1}{c}{T$_{eff}$-ZAHB}&
%\multicolumn{1}{c}{T$_{eff}$-col.-rel.}&
%\multicolumn{1}{c}{log(L/L$_{\odot}$)}&
%\multicolumn{1}{c}{log(M/M$_{\odot}$)}
\hline
\noalign{\smallskip}
NGC\,6397  & 30511& T183-1& 17:40:20.057& $-$53:41:42.91& 13.944& 13.866& 13.858&21.1 & 8.5\\
         &        &  T183-2     &             &               &        &       &       &20.2 &\\
      & 53234& T191& 17:40:56.022& $-$53:35:07.39& 14.037& 13.969& 13.923& 21.5 & 10.0\\
      & 52839& T193& 17:40:57.460& $-$53:35:32.79& 14.014& 13.954& 13.922& 24.0 &8.0 \\
\hline
NGC\,6752  & 22300& B652 & 19:11:31.490& $-$59:57:52.48& 14.337& 14.699& 14.783& $-$24.6 & 8.0\\
        & 28150& B2151-1& 19:11:01.965& $-$59:55:30.90& 14.202& 14.468& 14.554& $-$33.0 &2.5\\
        &   &   B2151-2     &             &      &         &               &       &$-$32.6&    \\
        & 22949& B2206& 19:11:00.931& $-$59:57:42.22& 14.202& 14.459& 14.532& $-$26.3 &2.5\\
\hline
\noalign{\smallskip}
\end{tabular}
\end{table*}

High resolution spectra of selected three BHB stars in NGC\,6397 (T183, T191, and T193) 
and three BHB stars in NGC\,6752 (B665, B2151, and B2206)
were observed on August 2 and 3, 2008 at ESO UVES at UT2. We used the UVES Dichroic 2 non-standard setting (437+700) to cover the  ranges
$\lambda\lambda$\,3758$-$4984, 5074$-$6930, and 7045-8904\,\AA{}. 
For the star T183, the slit width was set to 0.3\arcsec{} for the red arm, corresponding to a resolving power 
of $\sim$115\,000. The stars T191, T193, B652, B2151, and B2206  were observed with a slit 
width of 0.4\arcsec{} fot the red arm (R$\approx$95\,000).
For the blue arm, for all stars, we used a slit width of 0.4\arcsec{} to achieve a resolving power 
of $\sim$80\,000. 
%The resolving power is $\sim$80000  at 4500\,AA{} and $\sim$100000 at 6000 and 8000\,\AA{}.
%The spectra were reduced by the UVES pipeline Data Reduction Software 
%(version 2.5; Ballester et al.\ \cite{Ballester2000}).
The spectra were reduced by the UVES pipeline Data 
Reduction Software (version 4.3.0, the general description can be found in e.g. 
Ballester et al.\ \cite{Ballester2000}) and using standard IRAF routines. 
The S/N of the obtained UVES spectra range is of the order 75-125 in the near 
ultraviolet (UV) (3800\,\AA), 95-120 at  4500\AA{}, and 60-110 in the near-IR region 
containing the \ion{Ca}{ii} triplet lines. 
%In the spectral ranges $\lambda\lambda$ 5074-5550\,\AA{} and 7570-7686\,\AA{} the spectra have 
%very low S/N ratio and were not used in our analysis.
%The signal-to-noise ratios of the resulting UVES spectra range from 80 to 100 per 
%pixel in the one-dimensional spectrum. 
The stars T183 and B2151 were observed twice to examine potential spectrum variability,
like presence of chemical spots on the stellar surface or radial velocity variations due 
to the presence of a close companion.
The numbers assigned to the stars in NGC\,6397 and NGC\,6752 follow
Antony-Twarog \& Twarog (\cite{AntonyTwarog2000}) and Buonanno et al.\ (\cite{Buonanno1986}),
respectively. 
The observed stars are listed in Table~\ref{tab:photometry}. The rotational velocity \vsini{} was derived from 
the comparison
of observed and computed profiles, in particular using the line \ion{Mg}{ii} 4481.

%\begin{table*}
%\caption{The observed objects. The star numbers in NGC\,6752 refer to Buonanno
%et al.\ (\cite{Buonanno1986}), the star numbers in NGC\,6397 refer to Anthony-Twarog \& Twarog\ 
%(\cite{AntonyTwarog2000}). $v_{\rm shift}$ is obtained from the superposition of observed 
%spectra with the synthetic ones. It stands for the correction in the radial velocity and heliocentric 
%velocity.} 
%\label{tab:objects}
%\centering
%\begin{tabular}{rlrrrrr}
%\hline
%\hline
%\noalign{\smallskip}
%\multicolumn{1}{c}{Cluster}&
%\multicolumn{1}{c}{Star} &
%\multicolumn{1}{c}{RA (J2000)}&
%\multicolumn{1}{c}{Dec (J2000)}&
%\multicolumn{1}{c}{V} &
%\multicolumn{1}{c}{$v_{\rm shift}$} &
%\multicolumn{1}{c}{\vsini{}} \\
% & & & & & 
%\multicolumn{1}{c}{[km\,s$^{-1}$]} &
%\multicolumn{1}{c}{[km\,s$^{-1}$]} \\
%\hline
%\noalign{\smallskip}
%NGC\,6397&T183(ROB\,470) & 17:40:20.08 & $-$53:41:43.1 & 13.881 & 38.5 & 8.5\\
%        &T191            &17:40:55.99 & $-$53:35:07.5&13.954   & 39.5 & 10.0\\
%        &T193 (ROB\,56)  &  17:40:57.43 & $-$53:35:32.8& 13.963 & 43.0 &8.0 \\
%NGC\,6752& B652          &19:11:31.40 & $-$59:57:52.4 & 14.70 & $-$11.5 & 8.0\\
%        & B2151         &19:11:01.95 & $-$59:55:30.5 & 14.45 & $-$20.5 &2.5\\
%        & B2206         &19:11:00.93 & $-$59:57:42.0 & 14.30 & $-$13.5 &2.5\\
%\hline
%\noalign{\smallskip}
%\end{tabular}
%\end{table*}
\begin{table}
\caption{Cluster parameters} 
\label{tab:cluster_params}
\begin{flushleft}
\begin{tabular}{rlrlllrrrrr}
\hline
\hline
\noalign{\smallskip}
\multicolumn{1}{c}{NGC} & 
\multicolumn{1}{c}{[Fe/H]} &
\multicolumn{1}{c}{(m-M)$_{V}$} &
\multicolumn{1}{c}{E(B$-$V)}& 
\multicolumn{1}{c}{E(b$-$y)} 
 \\
\hline
\noalign{\smallskip}
6397 & $-$1.91$^{1,2}$& 12.19$^{1}$ & 0.179$^{3}$ &0.127$^{3}$  \\
          &                &             &$\pm$0.003&$\pm$0.002\\
6752 & $-$1.54$^{1}$ & 12.96$^{1}$ & 0.04$^{1}$                            \\
          & $-$1.61$^{2}$ &      & 0.05$^{2}$          &      &           \\
\hline
\noalign{\smallskip}
\end{tabular}
\end{flushleft}
$^{1}$ Zinn\ (\cite{Zinn1985}),\ $^{2}$ Grundhal et al.\ (\cite{Grundahl1999}),\\
$^{3}$ Anthony-Twarog\& Twarog (\cite{AntonyTwarog2000}).
%\end{flushleft}
\end{table}
%$^{1}$ Zinn (1985); $^{2}$ Grundhal et al. (1999);\\
%$^{3}$ Anthony-Twarog\& Twarog (2000).
%\end{table}

%\begin{table*}
%\caption{Cluster parameters} 
%\label{tab:cluster_params}
%%\begin{flushleft}
%\centering
%\begin{tabular}{rlrlllrrrrr}
%\hline
%\hline
%\noalign{\smallskip}
%\multicolumn{1}{c}{Cluster} & 
%\multicolumn{1}{c}{[Fe/H]} &
%\multicolumn{1}{c}{(m-M)$_{V}$} &
%\multicolumn{1}{c}{E(B$-$V)}& 
%\multicolumn{1}{c}{E(b$-$y)} 
% \\
%\hline
%\noalign{\smallskip}
%NGC\,6397 & $-$1.91$^{1,2}$& 12.19$^{1}$ & 0.179$\pm$0.003$^{3}$ &0.127 $\pm$0.002$^{3}$ \\
%NGC\,6752 & $-$1.54$^{1}$ & 12.96$^{1}$ & 0.04$^{1}$                            \\
%          & $-$1.61$^{2}$ &      & 0.05$^{2}$          &      &           \\
%\hline
%\noalign{\smallskip}
%\end{tabular}
%\begin{flushleft}
%$^{1}$ Zinn\ (\cite{Zinn1985}),\ $^{2}$ Grundhal et al.\ (\cite{Grundahl1999}),\
%$^{3}$ Anthony-Twarog\& Twarog (\cite{AntonyTwarog2000}).
%\end{flushleft}
%\end{table*}

\section{Abundance analysis}

A standard model atmosphere analysis was made using
as starting models those computed  
with the ATLAS9 code (Kurucz \cite{Kurucz1993}) assuming a solar chemical composition
and as final models those computed with the ATLAS12 code 
(Kurucz \cite{Kurucz2005}) for individual stellar abundances. We used Linux versions
of the codes as described in Shordone et al. (\cite{shord2004}) and Castelli (\cite{Castelli2005a}).

\subsection{Atmospheric parameters}

The gravity was computed using its definition:

\begin{eqnarray}
 g=g_{\odot} \frac{M}{R^2} \nonumber
%\begin{displaymath}  g=g_{\odot}(M/M_{\odot})/(R/R_{\odot})^2\end{displaymath}
\end{eqnarray}
\noindent
where $g_{\odot}$ is the surface gravity of the sun and the mass, $M$, and radius, $R$, of the star are in solar units.
%are the mass and radius of the star.
According to the relation of radius to luminosity and therefore to  
bolometric magnitude, the gravity is given by:

\begin{eqnarray}
 \log g & = & \log g_{\odot} + \log (M/M_{\odot}) \nonumber \\
        & + & 0.4(M_{\rm bol}-M_{{\rm bol},\odot}) \nonumber \\
        & + & 4 \log (T_{\rm eff}/T_{\odot}) \nonumber
\end{eqnarray}

\noindent
where the  bolometric magnitude is:

\begin{eqnarray}
 M_{\rm bol}=V-(m-M)_{V}+BC_{V} \nonumber
\end{eqnarray}

\noindent
and  (m-M)$_{V}$ is the distance modulus, while V is the visual magnitude
corrected for reddening. We adopted  A(V)=3.1E(B$-$V) and 
distance modulus and reddening listed in Table~\ref{tab:cluster_params}. 
Further we assumed M$_{{\rm bol}, \odot}$=4.74 (Bessell et al.\ \cite{Bessell1998}) and used 
the bolometric corrections BC$_{V}$  computed by Bessell et al.\ (\cite{Bessell1998}) 
for solar metallicity atmospheric models having zero microturbulent velocity.
In fact, all the studied stars have [Fe/H] nearly solar or even super-solar. 
As stellar total
mass we assumed M/M$_{\odot}$=0.6. Behr (\cite{Behr2003}) estimated that an uncertainty of
 0.06M$_{\odot}$,  as due to the uncertainties in
temperature and metallicity,  would produce an uncertainty  of 0.04\,dex in \logg{}. 
Under the assumption of solar parameters \teff{}$_{, \odot}$=5777\,K, \logg{}$_{\odot}$=4.4377 we computed 
stellar gravities using the equation presented above.
Because \teff{} is not known in advance, 
we adopted as \logg{} the averages of the gravities obtained for 
\teff{}=11\,000\,K, 11\,500\,K, 12\,000\,K, and 12\,500\,K.
%Because \teff{} is not known in advance, we adopted as \logg{} the average 
%of the gravities obtained from the above relation for
%\teff{}=11\,000\,K, 11\,500\,K, 12\,000\,K, and 12\,500\,K.
%Under the assumption of solar parameters \teff{}$_{, \odot}$=5777\,K, \logg{}$_{\odot}$=4.4377 we computed 
%stellar gravities used the equation presented above. 
They are presented in Table~\ref{tab:gravity_det}.  
For B652 we computed the gravities obtained from two different determinations
of the visual magnitude, i.e.\ V=14.70 from Buonanno et al.\ (\cite{Buonanno1986}) and V=14.743
as quoted by Moehler et al.\ (\cite{Moehler2000}). The corresponding maximum gravity 
difference amounts to 0.02\,dex. 
An estimated error of 0.02\,mag in E(B-V) affects the gravity 
by $\pm$0.03\,dex. However, the error of 0.003\,dex as estimated by
Antony-Twarog \& Twarog (2000) induces a gravity error of only 0.004\,dex.
An error of 0.07\,mag in distance modulus corresponding to
about 3.5\% in the distance (Gratton et al.\ \cite{Gratton2003}) leads to an 
error of $\pm$0.03\,dex in the gravity.

\begin{table*}
\caption{The gravity determination.} 
\label{tab:gravity_det}
\centering
\begin{tabular}{lccccccc}
\hline
\noalign{\smallskip}
   & T183  &  T191  & T193 &\multicolumn{2}{c}{B652} & B2151 & B2206\\
 \hline
\noalign{\smallskip}
\multicolumn{1}{r}{V} & 13.881& 13.954& 13.963& 14.70 & 14.743          & 14.45 & 14.30\\ 
\hline
\noalign{\smallskip}
\teff{} & \multicolumn{7}{c}{\logg}\\
\hline
\noalign{\smallskip}
11000   &  3.71 &  3.74 &  3.74 & 3.89 & 3.91 &3.79   & 3.73 \\
11500   &  3.75 &  3.78 &  3.78 & 3.93 & 3.94 &3.83   & 3.77\\
12000   &  3.78 &  3.81 &  3.81 & 3.96 & 3.98 &3.86   & 3.80\\
12500   &  3.81 &  3.84 &  3.85 & 3.99 & 4.01 &3.89   & 3.83\\
average \logg: & 3.76$\pm$0.04&3.79$\pm$0.04&3.79$\pm$0.04&3.94$\pm$0.04&3.96$\pm$0.04& 3.84$\pm$0.04&3.78$\pm$0.04\\
\hline
\noalign{\smallskip}
\end{tabular}
\end{table*}

Once the gravity was fixed for each star, we derived the temperature  
by forcing \ion{Fe}{i} and \ion{Fe}{ii} abundances to be consistent. 
The resulting temperatures are given in Table~\ref{tab:teff_logg}. 
The analyzed \ion{Fe}{i} and \ion{Fe}{ii} lines are listed in Table~\ref{tab:analyzed_lines} without any asterisk.
We estimated that an uncertainty of 0.2\,dex in the gravity corresponds 
to a temperature error of 300\,K which produces an uncertainty of 0.1\,dex
 in the iron abundance.
To estimate the temperature error related to a different choice of 
 \ion{Fe}{i} and \ion{Fe}{ii} we measured the equivalent widths for a few 
other lines. They are marked with an asterisk in the Table~\ref{tab:analyzed_lines}. The extended 
line number modifies the temperature by 100\,K.

\begin{table*}
\caption{\teff{} derived from the \ion{Fe}{i} and 
\ion{Fe}{ii} ionization equilibrium for a fixed gravity.
Results are shown both from ATLAS9 models with [M/H]=0.0 
and ATLAS12 models computed for individual abundances.
The presented abundances $\log\epsilon$(Fe\,I) and $\log\epsilon$(Fe\,II) are defined as 
$\log$(N$_{\rm elem}$/N$_{\rm tot}$).
} 
\label{tab:teff_logg}
%\begin{flushleft}
\centering
\begin{tabular}{lr|rrr|rrr|rlc}
\hline
\hline
\noalign{\smallskip}
\multicolumn{1}{c}{Star} & 
\multicolumn{1}{c}{$\log\,g$} & 
\multicolumn{1}{c}{\teff{}}& 
\multicolumn{1}{c}{$\log\epsilon$(Fe\,I)}&
\multicolumn{1}{c}{$\log\epsilon$(Fe\,II)}&
\multicolumn{1}{c}{\teff{}}& 
\multicolumn{1}{c}{$\log\epsilon$(Fe\,I)}&
\multicolumn{1}{c}{$\log\epsilon$(Fe\,II)}&
\multicolumn{1}{c}{\teff{}}&
\multicolumn{1}{c}{\logg}&
\multicolumn{1}{c}{Source} \\
\hline
\noalign{\smallskip}
 & &
\multicolumn{3}{c}{ATLAS9} &
\multicolumn{3}{c}{ATLAS12} &
\multicolumn{3}{c}{Literature}\\
\hline
\noalign{\smallskip}
T183  & 3.75 &11480 &$-$4.18$\pm$0.09 &$-$4.18 $\pm$0.11 & 11390 &  $-$4.19$\pm$0.09 &$-$4.19$\pm$0.11&11950$\pm$200 & 3.59 & 1\\
T191  & 3.80 &11770 &$-$4.68$\pm$0.06 &$-$4.68 $\pm$0.09 & 11630 & $-$4.70$\pm$0.06 &$-$4.69$\pm$0.10\\
T193  & 3.80 &11480 &$-$4.27$\pm$0.09 &$-$4.27 $\pm$0.10 & 11370 & $-$4.29$\pm$0.09 &$-$4.29$\pm$0.10&11430$\pm$190 & 3.56 & 1\\
B652  & 3.95 &12250 &$-$3.80$\pm$0.11 &$-$3.80 $\pm$0.09 & 11960 & $-$3.86$\pm$0.11 &$-$3.85$\pm$0.09&12500$\pm$230 & 3.98$\pm$0.07 &2\\
B2151 & 3.85 &11400 &$-$4.18$\pm$0.05 &$-$4.18 $\pm$0.11 & 11310 & $-$4.20$\pm$0.05 &$-$4.20$\pm$0.11\\
B2206 & 3.80 &11290 &$-$4.21$\pm$0.05 & $-$4.21 $\pm$0.08 &11200 & $-$4.22$\pm$0.05 &$-$4.22$\pm$0.08  \\
\hline
\noalign{\smallskip}
\end{tabular}
%\end{flushleft}
\begin{flushleft}
(1) de Boer et al.\ (\cite{deBoer1995}),
(2) Moehler et al.\ (\cite{Moehler2000}) for [M/H]=0.0.
\end{flushleft}
\end{table*}

For B-type stars also Balmer profiles can be used to derive the atmospheric
parameters, in particular the gravity because Balmer lines
 are rather  insensitive to the temperature for \teff{}$>$ 11\,000\,K. 
Using a $\chi^{2}$ method we fitted the observed 
H$\alpha$, H$\beta$, H$\gamma$, and H$\delta$ profiles, normalized to the 
continuum level and shifted to the laboratory wavelength, to the grid of 
Balmer profiles based  on ATLAS9 model atmospheres computed for [M/H]=0.0 
and microturbulent velocity  $\xi$=0.0\,km\,s$^{-1}$. 
Balmer lines were computed with the BALMER9 code (Kurucz \cite{Kurucz1993}). 
Also in this case, we adopted solar metallicity rather than cluster 
metallicities ($-$1.91 for NGC\,6397 and
$-$1.61 for NGC\,6752) due to the nearly solar or even 
super-solar [Fe/H] abundances of the studied stars.
The parameters from Balmer profiles are listed in
Table~\ref{tab:balmer_profiles}. The average temperatures and gravities are smaller than those
listed in Table~\ref{tab:teff_logg}. The large standard deviations  associated 
to the average temperatures are partly due to the low sensitivity of 
the Balmer profiles to temperature and partly due to
the difficulty in fixing the continuum level over the observed Balmer 
profiles caused by spectral imperfection in the UVES spectra related 
to the echelle orders.
While the order combining does not affect much the analysis of most of the lines, 
it causes a challenge for Balmer lines in that their extended wings
span over more than a single order. 
The discrepancy in the parameters derived with the two methods can also
be ascribed to the solar metallicity used for computing Balmer profiles,
while for almost all elements the abundances are different from the solar ones.
In fact we tested that, for the same parameters,
the Balmer lines from ATLAS12 models with individual abundances are 
weaker than those computed from  ATLAS9 models with solar metallicity, 
so that ATLAS12 models would
require a higher gravity and/or a lower \teff{} to fit the observed Balmer line profiles.

Since the first method is almost model independent as far as the
gravity is concerned, we adopted 
as model parameters those presented in Table~\ref{tab:teff_logg}.
The zero microturbulent velocity agrees with the value usually
adopted for field B-type peculiar stars. We did not find any indication that a
higher value of $\xi$ would have been a better choice.

\begin{table*}
\caption{Stellar parameters from Balmer profiles.} 
\label{tab:balmer_profiles}
\centering
\begin{tabular}{lcccccccccc}
\hline
\noalign{\smallskip}
\multicolumn{1}{c}{Star} & 
\multicolumn{2}{c}{H$_{\alpha}$} & 
\multicolumn{2}{c}{H$_{\beta}$}& 
\multicolumn{2}{c}{H$_{\gamma}$}&
\multicolumn{2}{c}{H$_{\delta}$}&
\multicolumn{2}{c}{Average} \\
\hline
 &
\multicolumn{1}{c}{\teff{}} &
\multicolumn{1}{c}{\logg{}} &
\multicolumn{1}{c}{\teff{}} &
\multicolumn{1}{c}{\logg{}} &
\multicolumn{1}{c}{\teff{}} &
\multicolumn{1}{c}{\logg{}} &
\multicolumn{1}{c}{\teff{}} &
\multicolumn{1}{c}{\logg{}} &
\multicolumn{1}{c}{\teff{}} &
\multicolumn{1}{c}{\logg{}} \\
\hline
\noalign{\smallskip}
T~183& 11300 & 3.9 & 10850 & 3.5 & 10750 & 3.6 & 11650 & 4.0  & 11137$\pm$361 & 3.75$\pm$0.21\\
T~191& 10350 & 3.4  &12000 & 4.1 & 11750 & 3.9 & 11750 & 4.0  & 11462$\pm$650 & 3.85$\pm$0.27 \\
T~193& 10400 & 3.4 & 11200 & 3.5 & 10950 & 3.6 & 10500 & 3.6  & 10763$\pm$327 & 3.52$\pm$0.08\\
B~652& 10750 & 3.6 & 11300 & 3.4 & 11000 & 3.6 & 10900 & 3.5  & 10988$\pm$201 & 3.52$\pm$0.08 \\
B2151 & 10450 & 3.6 & 11150 & 3.5 & 10500 & 3.6 & 10500 & 3.6  &10650$\pm$289 & 3.57$\pm$0.04 \\
B2206 & 10450 & 3.6 & 10900 & 3.5 & 10650 & 3.6 & 10500 & 3.6  & 10625$\pm$175 & 3.57$\pm$0.04 \\
\hline
\noalign{\smallskip}
\end{tabular}
\end{table*}

\subsection{Individual abundances in the stars}

Spectral lines belonging to elements \ion{He}{i}, \ion{Mg}{ii}, \ion{P}{ii}, \ion{Ca}{ii},
\ion{Ti}{ii}, \ion{Cr}{ii}, \ion{Mn}{ii}, \ion{Fe}{i}, \ion{Fe}{ii},
\ion{Ni}{ii}, and \ion{Y}{ii} have been detected in all stars,
except for \ion{P}{ii} in T191 where instead lines of \ion{O}{i},
\ion{Si}{ii}, and \ion{Co}{ii} were detected.
\ion{Si}{ii} lines were also observed in T183, while \ion{Na}{i} was
observed in B652 and B2206.

Equivalent widths were measured for \ion{Ti}{ii}, \ion{Cr}{ii},
\ion{Fe}{i}, and \ion{Fe}{ii} by integrating the residual intensities 
over the profiles. Also equivalent widths were measured for 
a few lines of \ion{P}{ii} and \ion{Y}{ii} which were 
not too weak or not too much affected
by noise. For these lines the abundances were obtained with a Linux version
(Castelli \cite {Castelli2005b})
of the WIDTH code (Kurucz \cite {Kurucz1993}).
For weak lines with no measurable equivalent widths
we derived the abundance from the line profiles with the synthetic
spectrum method. Synthetic spectra were computed with a Linux version
(Sbordone et al. \cite {shord2004}) of
the SYNTHE code (Kurucz \cite{Kurucz2005}).
When no lines were observed for a given element an upper abundance limit 
was fixed by reducing the intensity of the computed line at the level
of the noise.
The lines used to derive abundances are listed in Table~\ref{tab:analyzed_lines}.
Mean abundances for each species based on ATLAS9 model
atmospheres are summarized in Table~\ref{tab:ATLAS9abundances}. 
In this table errors are given only for lines with measured 
equivalent widths, and no errors are presented for lines
with abundances from comparison of observed and computed profiles
using synthetic spectrum method. For \ion{Mn}{ii} we did not use equivalent
widths because Mn is affected by hyperfine structure. When available, the hyperfine components are
considered in the synthetic spectrum.

\begin{table*}
\caption{Abundances from ATLAS9 models computed for [M/H]=0.0
and microturbulent velocity $\xi$=0\,km\,sec$^{1}$. At this step
the helium abundance was assumed solar. 
%Errors are given only for lines with measured 
%equivalent widths, and no errors are presented for lines with abundances from comparison of observed and computed profiles
%using synthetic spectrum method. For \ion{Mn}{i} we did not use equivalent
%widths because Mn is affected by hyperfine structure. The hyperfine components are
%considered in the synthetic spectrum,
} 
\label{tab:ATLAS9abundances}
\centering
\begin{tabular}{rrrrrrrrrr}
\hline
\noalign{\smallskip}
\multicolumn{1}{c}{Elem.} & 
\multicolumn{1}{c}{T183}& 
\multicolumn{1}{c}{T191} & 
\multicolumn{1}{c}{T193}&
\multicolumn{1}{c}{NGC\,6397}&
\multicolumn{1}{c}{B652}&
\multicolumn{1}{c}{B2151}&
\multicolumn{1}{c}{B2206}&
\multicolumn{1}{c}{NGC\,6752}&
\multicolumn{1}{c}{Sun}
  \\
\hline
\noalign{\smallskip}
  &(11480,3.75) &(11770,3.8) &(11480,3.80)& &(12250,3.95) &(11400,3.85) &(11290,3.80) & & \\
\hline
\noalign{\smallskip}
\ion{C}{ii} &$\le$$-$5.02      &$\le$$-$5.02 &$\le$$-$5.02 &$-$5.43&$\le$$-$4.52 &$\le$$-$4.52 &$\le$$-$4.52&$-$5.13 & $-$3.52 \\
\ion{N}{i} &$\le$$-$5.12      &$\le$$-$5.12 &$\le$$-$5.12 & $-$6.03&$\le$$-$5.12 &$\le$$-$5.82 &$\le$$-$5.82&$-$5.73& $-$4.12\\
\ion{O}{i}  &$\le$$-$5.21      &$-$4.80 &$\le$$-$5.51 &$-$5.12&$\le$$-$5.51 &$\le$$-$5.51 &$\le$$-$5.51 &$-$4.82& $-$3.21 \\
\ion{Na}{i} &$\le$$-$6.71      &$\le$$-$6.71 &$\le$$-$6.71 &$-$7.62&$-$5.90 &$\le$$-$6.71 &$-$6.45 &$-$7.32& $-$5.71 \\ 
\ion{Mg}{ii} & $-$6.11          &$-$6.38 & $-$6.00 &$-$6.37&$-$5.68 &$-$5.75 &$-$5.71 &$-$6.07& $-$4.46 \\
\ion{Al}{i} &$\le$$-$6.75      &$\le$$-$6.75 &$\le$$-$6.75 &$-$7.48&$\le$$-$7.15 &$\le$$-$7.25 &$\le$$-$7.00 & $-$7.18& $-$5.57 \\
\ion{Si}{ii} &$-$7.49           &$-$6.76 &$\le$$-$7.80 &$-$6.40&$\le$$-$7.60 &$\le$$-$7.70 &$\le$$-$8.00 & $-$6.10&$-$4.49\\
\ion{P}{ii} &$-$5.40           &$\le$$-$6.59 &$-$5.43$\pm$0.10 &$-$8.50&$-$4.89$\pm$0.25 &$-$5.53 &$-$5.60$\pm$0.10 &$-$8.20& $-$6.59\\
\ion{S}{ii}  &$\le$$-$6.21      &$\le$$-$6.41 &$\le$$-$6.21 &$-$6.62&$\le$$-$6.21 &$\le$$-$6.21 &$\le$$-$6.21 &$-$6.32& $-$4.71\\
\ion{Ca}{ii} &$-$5.88           &$-$5.53 &$-$6.22 &$-$7.59&$-$5.85 &$-$6.34 &$-$6.28 &$-$7.29 & $-$5.68 \\
\ion{Sc}{ii} &$\le$$-$9.87      &$\le$$-$9.87 &$\le$$-$10.37 &$-$10.78 &$\le$$-$9.87 &$\le$$-$10.37 &$\le$$-$10.37 &$-$10.48& $-$8.87 \\
\ion{Ti}{ii} & $-$6.34$\pm$0.09 &$-$5.31$\pm$0.11 & $-$6.69$\pm$0.09 &$-$8.93&$-$7.00$\pm$0.11 &$-$6.87$\pm$0.07 &$-$6.88$\pm$0.07 &$-$8.63&$-$7.02\\
%\ion{V}{ii}  &$\le$$-$8.54      &$\le$$-$9.54  &$\le$$-$8.54 &$-$9.95&$\le$$-$9.04 &$\le$$-$9.54 &$\le$$-$9.54 &$-$9.65& $-$8.04 \\  
\ion{Cr}{ii} &$-$7.05$\pm$0.05 &$-$6.43$\pm$0.03 &$-$6.54$\pm$0.05 &$-$8.28&$-$6.81$\pm$0.04 &$-$6.88$\pm$0.03 &$-$6.80$\pm$0.05 &$-$7.98&$-$6.37\\
\ion{Mn}{ii} &$-$6.05           &$-$5.35 &$-$5.44 &$-$8.56&$-$5.75 &$-$6.50 &$-$6.30 &$-$8.26&$-$6.65\\
\ion{Fe}{i} & $-$4.17$\pm$0.09  & $-$4.68$\pm$0.06 & $-$4.27$\pm$0.09 &$-$6.45& $-$3.80$\pm$0.11 & $-$4.18$\pm$0.05 & $-$4.21$\pm$0.05 &$-$6.15&$-$4.54\\
\ion{Fe}{ii}&$-$4.17 $\pm$0.11 &$-$4.68$\pm$0.09&$-$4.27$\pm$0.10&$-$6.45&$-$3.80$\pm$0.09&$-$4.18$\pm$0.11&$-$4.21$\pm$0.08&$-$6.15&$-$4.54\\
\ion{Co}{ii} &$\le$$-$7.92 & $-$6.12  &$\le$$-$7.12&$-$9.03&$-$&$\le$$-$7.92&$\le$$-$7.92&$-$8.73&$-$7.12\\
\ion{Ni}{ii}&$-$6.14& $-$6.00 &$-$6.25 &$-$7.70&$-$6.30 &$-$6.60 &$-$6.60 &$-$7.40& $-$5.79\\
\ion{Sr}{ii} & $\le$$-$10.57&$\le$$-$10.07 &$\le$$-$10.07 &$-$10.98&$\le$$-$10.07 &$\le$$-$10.57 &$\le$$-$10.57 &$-$10.68& $-$9.07\\
\ion{Y}{ii} & $-$8.20&$-$8.20 &$-$8.15$\pm$0.05 &$-$11.71&$-$7.43$\pm$0.12 &$-$8.50 &$-$8.34$\pm$0.08 &$-$11.41& $-$9.80\\
\ion{Zr}{ii}& $-$11.35 &$-$11.35 &$-$11.35 &$-$11.35&$-$11.05&$-$11.05&$-$9.01&$-$11.05&$-$9.44\\
\hline
\noalign{\smallskip}
\end{tabular}
\end{table*}

ATLAS9 models were not used to derive the He abundance.
As a matter of fact, inaccurate result are expected due to the inconsistency between the populations 
from the ATLAS9 model computed for a solar He/H ratio and the populations
obtained by SYNTHE for modified He/H values.
% gives inaccurate results. 
The helium abundance was derived later by comparing the observed
\ion{He}{i} profiles with 
profiles computed from an ATLAS12 models used for the individual
stellar abundances. 
%He I profiles with profiles based on ATLAS12 models.
%...the helium abundance was derived FOR A SECOND TIME-
%The SYNTHE code was not used to compute helium abundances
%based on  ATLAS9 models. As a matter of fact, the inconsistency between the populations 
%from the ATLAS9 model computed for a solar He/H ratio and the populations
%obtained by SYNTHE for modified He/H values gives inaccurate results. 
%The helium abundance was derived
%for a second time by comparing the observed \ion{He}{i} profiles with 
%profiles computed from an ATLAS12 models computed for the individual
%stellar abundances. 

We applied an iterative procedure. For each star, an ATLAS12 model was computed adopting the ATLAS9 parameters listed
in Table~\ref{tab:teff_logg} and the abundances listed
in Table~\ref{tab:ATLAS9abundances} for elements having observed lines in the spectra.
For elements without observed lines the abundance was assumed
to be equal to the metallicity of the cluster provided that
the abundance limit given in Table~\ref{tab:ATLAS9abundances}
%Table~\ref{tab:ATLAS12abundances}  which presents ATLAS12 abundances, 
exceeds the cluster metallicity. 
Otherwise, as for oxygen and silicon, the abundance limit was adopted.
New abundances from equivalent
widths and line profiles were determined with ATLAS12.
Keeping fixed the model independent gravity, a new \teff{} was determined from the condition 
that the \ion{Fe}{i} and \ion{Fe}{ii}
abundances concur. The iteration was continued until ATLAS12 abundances and
abundances derived from equivalent widths and line profiles agreed within
0.01\,dex. 

In Table~\ref{tab:teff_logg}, in Cols.~6, 7 and 8, we present the ATLAS12 final  parameters and 
the corresponding iron abundances.
In Table~\ref{tab:ATLAS12abundances} we list the final abundances used for computing the ATLAS12 models.
For elements not listed in Table~\ref{tab:ATLAS12abundances} the
adopted abundance was that yielded by the cluster metallicity.
Finally, Table~\ref{tab:rel_abundances} summarizes the peculiar abundances 
relative to the solar values. 
%In all six studied BHB stars the majority of observed spectral lines belongs to the same elements:
%\ion{He}{i},  \ion{Mg}{ii},  \ion{P}{ii},  \ion{Ca}{ii},  \ion{Ti}{ii},  \ion{Cr}{ii},
%\ion{Mn}{ii}, \ion{Fe}{i}, \ion{Fe}{ii}, \ion{Ni}{ii}, and \ion{Y}{ii}.  

The He underabundance was for quite a long time the only 
anomaly measured in cluster BHB stars (Greenstein et al.\ \cite{green1967}).
This underabundance is generally attributed to gravitational settling.
Only the \ion{He}{i} lines at $\lambda\lambda$ 4026.2, 4471.5, 5875.7, 
and 6678.15\,\AA{} are well observable in all stars. 
The lines at $\lambda\lambda$ 3819.6, 
4120.8, and 4387.93\,\AA{} can be observed as broad weak absorptions, whereas
\ion{He}{i} lines at $\lambda\lambda$  3867.5, 4143.76, 4713.2, and 4921.93\,\AA{}  are either 
not observed or are components of blends with \ion{Fe}{i} and \ion{Fe}{ii} 
lines.
The \ion{He}{i} line data are those listed in Appendix\,A of Castelli \& Hubrig (\cite{CastelliHubrig2004b})
and  the computation of the line profiles is described in the same paper.
%Only \ion{He}{i} 5875.6\,\AA{} can't be reproduced with the abundance well 
%fitting all the other \ion{He}{i} lines. It requires an \ion{He}{i} abundance 0.2\,dex larger.
In all stars, except for B652,  all \ion{He}{i} lines can 
be well reproduced by the same abundance, so that there are no signs of 
helium  vertical abundance stratification. 
In B652  the observed \ion{He}{i} profile at 5875.7\,\AA{} 
requires an abundance 0.2\,dex larger than that 
derived from the other lines. 
We could explain this discrepancy as due either to 
NLTE effects or to vertical abundance stratification. But this last case
would imply an increasing of the abundance toward the upper atmospheric layers
in contrast with what happens in field stars that show shallow line cores
and extended wings.

Elements more abundant than the cluster metallicity  in all 
BHB stars are  {Ca}, {Ti}, {Cr}, {Mn}, {Fe},
{Ni}, and {Y}. Phosphorus is overabundant in all 
stars except for T191 where it was not observed.  
Na is overabundant in
both B652 and B2206, Co is overabundant only in T191 and Zr is overabundant in B2206.

On the other side, in addition to {He} which is always underabundant
also {Si} always appears 
less abundant  than the cluster metallicity (see e.g.\ Fig.~\ref{fig:si_fit}). 
{O} is also underabundant, except for T191, which shows oxygen overabundance by about 0.3\,dex.
The Si abundance can be best determined in T183 and T191 due to
the presence of weak spectral lines. 
Further, we note that {Al} is less abundant than cluster metallicity in B2151.
No definite differences were observed between the abundances of neutral and ionized states of studied 
elements indicating absence of vertical elemental stratifications in the atmospheres of BHB stars. 
We note, however, that apart from \ion{Fe}{i} lines, all other lines of neutral states like those of 
\ion{Cr}{i} and \ion{Mn}{i} are extremely weak and cannot be used for the abundance analysis.

\begin{figure}
\begin{center}
\includegraphics[width=0.45\textwidth,angle=0,clip=]{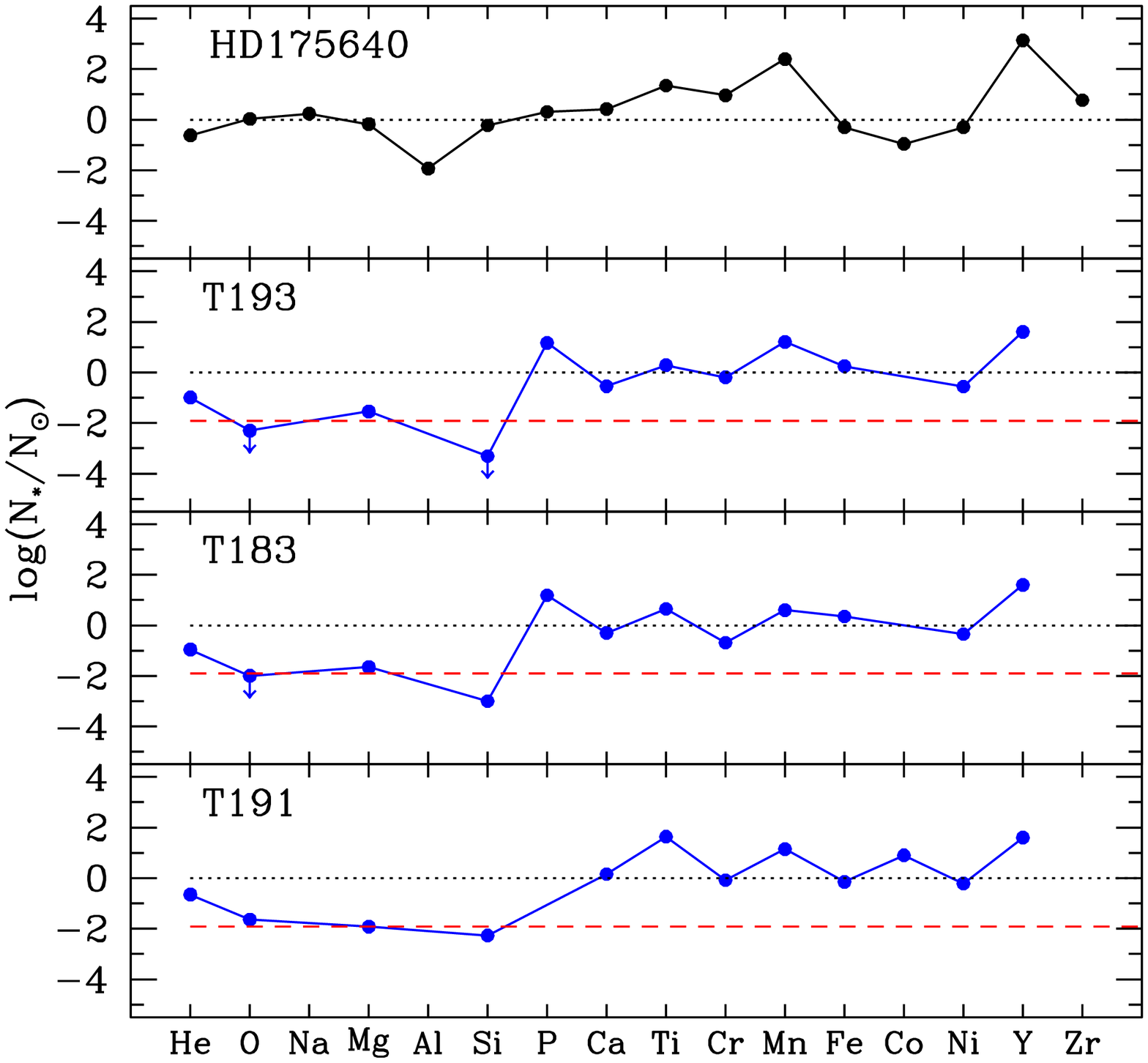}
\includegraphics[width=0.45\textwidth,angle=0,clip=]{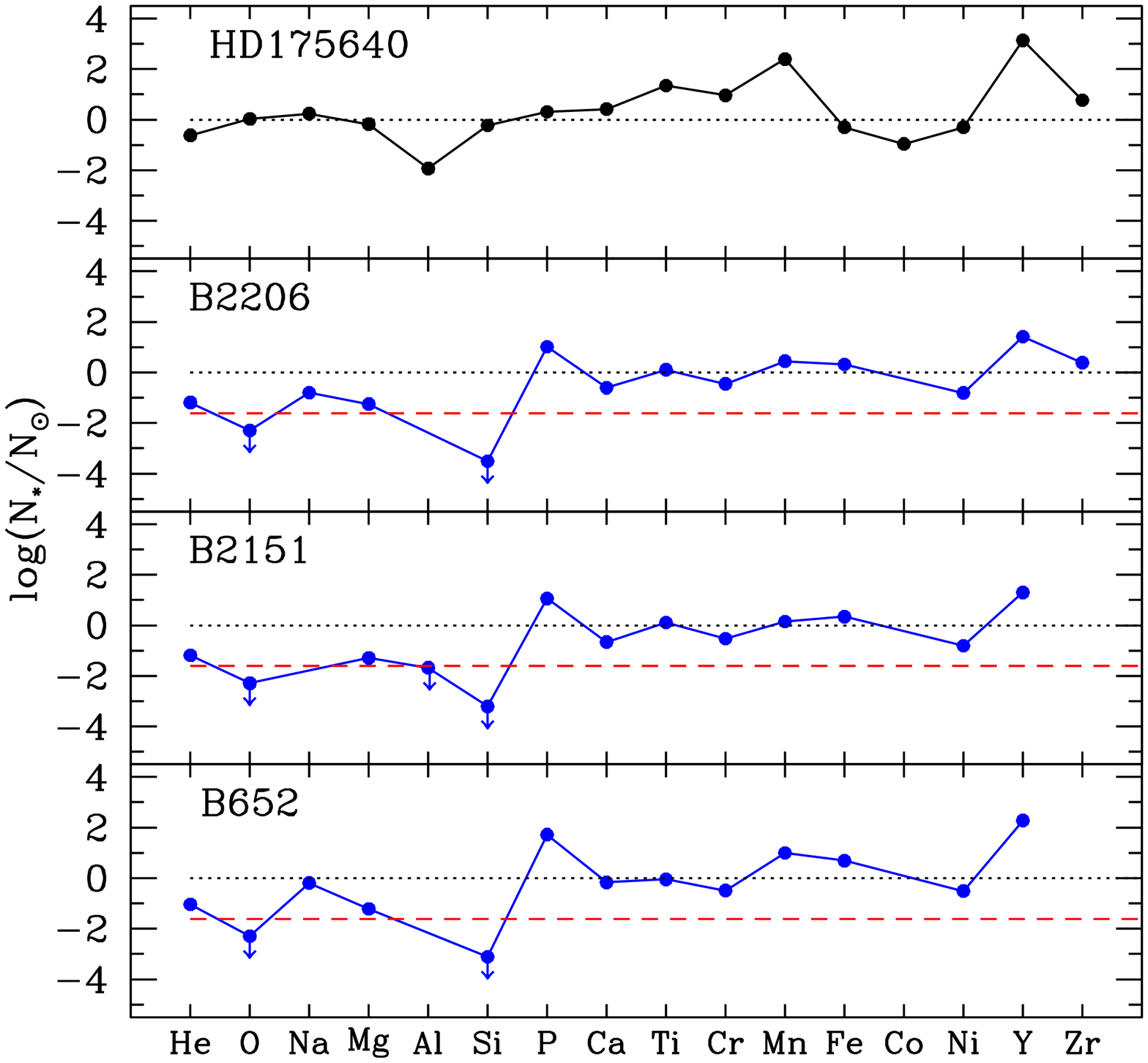}
\end{center}
\caption{
Final ATLAS12 abundances $\log$(N$_{\rm elem}$/N$_{\rm tot}$) for the members NGC\,6397 and 
of NGC\,6752 relative to the solar 
abundances (presented by the dotted horizontal line). Note that the error bars 
cannot be presented on these plots as they are smaller than the plotted symbols.
The dashed line indicates the cluster abundances.
%are too small to present them on these plots, they are hidden behind the symbols. 
BHB stars are shown in the order 
of increasing \teff from top to bottom. The elements for which only an upper limit of the 
abundance is available are indicated by a downward arrow. The upper panels present abundances of the typical 
HgMn star HD\,175640.}
\label{fig:abund_plots}
\end{figure}

In Fig.~\ref{fig:abund_plots} we present for each star final ATLAS12 abundances $\log$(N$_{\rm elem}$/N$_{\rm tot}$) relative to the solar 
abundances. The top panels present abundances of the typical 
HgMn star HD\,175640 of a similar temperature with \teff=12\,000\,K (Castelli \& Hubrig\ \cite{CastelliHubrig2004b}). 
The largest abundance dispersions within the cluster are found for elements {Si}, {P},
{Ti}, {Mn}, and {Co} in NGC\,6397. In the cluster NGC\,6752 the 
largest abundance dispersions are found for {Na}, {P}, {Mn}, {Y} and {Zr}.
Comparing the abundances between all studied stars we note that the star T191 clearly stands out in the sample due to 
the underabundance of phosphorus and strong overabundance of titanium and cobalt. As we show in Sect. 4.2 
it is also the only star which exhibits emissions in \ion{Ti}{ii} lines in the region 6000-6110\,\AA{}. 
This star should certainly be studied in more detail in future follow-up studies.
No evident trends of elemental abundances with fundamental parameters \teff{}, \logg{} and \vsini{} were found 
in our sample stars, although it is possible that the yttrium abundance is slightly increasing with 
\teff{}. Also the largest titanium overabundance was found in T191 which rotates slightly faster than 
other sample stars. On the other hand, the existence of trends should be further examined using 
observations of a larger 
sample of BHB stars, as our sample stars have a rather small range of parameters. 
The comparison of abundance patterns between BHB stars and the typical HgMn star HD\,175640 shows 
somewhat similar trends, apart from the low metallicity of the major part of the studied elements.
To our knowledge, no other abundance study of any hot BHB star in NGC\,6397 has been performed in the past.
The overabundance of iron and normal magnesium abundance were reported by Moehler et al.\ (\cite{moehler1999})
who studied 19 BHB stars in  NGC\,6752 using low-resolution spectra obtained with the ESO 1.52\,m telescope.
While the authors find an enrichment of iron by a factor of 50 on average with respect to cluster abundance 
and a magnesium abundance consistent with the cluster metallicity
in BHB stars with \teff{}$>$ 11\,000\,K,
our abundance results show an enrichment of iron by an average factor of 100 and 
enrichment of magnesium by a factor of $\sim$2, for both NGC\,6397 and NGC\,6752.
Only for T191 in NGC\,6397 the magnesium abundance is consistent with the cluster abundance.
These differences in the abundance analyses are possibly explained by the 
higher resolution of our spectra and advanced methods of abundance determinations using now 
available ATLAS12 models.
Glaspey et al.\ (\cite{glaspey1989}) studied in addition to iron and magnesium 
silicon and phosphorus in a somewhat hotter BHB star, CL1083, in NGC\,6397 with parameters 
\teff{}=16\,000\,K and \logg=4.0.
While silicon abundance was found consistent with the cluster abundance, no phosphorus lines 
could be identified in the spectrum of this star.
%Chemical composition of BHB stars in NGC\,6397 was previously studied by Lambert et al.\ (\cite{lambert1992})
%using low resolution spectra (R=18\,000) obtained at the CTIO 4\,m telescope.
%\changea{

\begin{figure}
\begin{center}
\includegraphics[width=0.45\textwidth,angle=0,clip=]{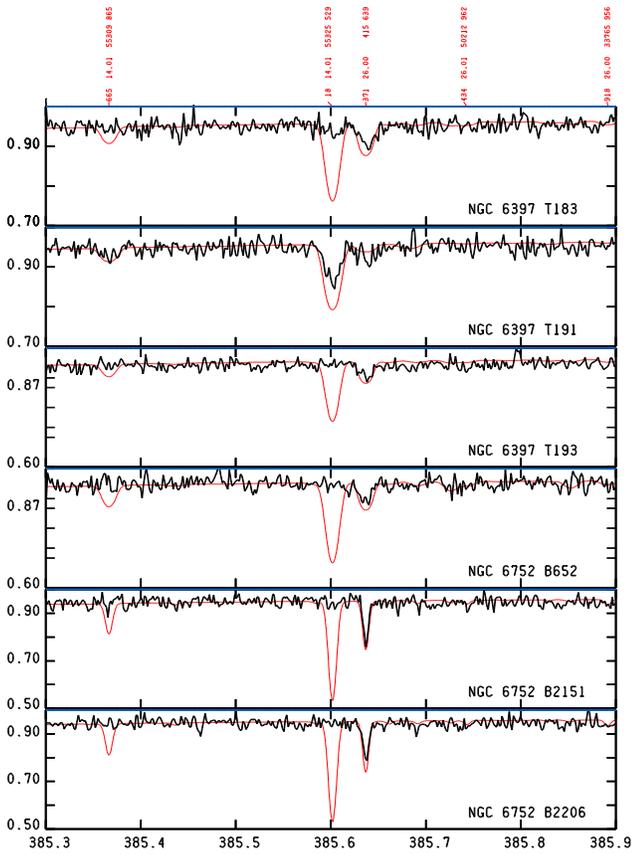}
\end{center}
\caption{
\ion{Si}{ii} 3856.18 computed with the
abundance of the clusters (thin line), i.e.\ $-$6.40 for NGC\,6397
and $-$6.10 for NGC\,6752.
}
\label{fig:si_fit}
\end{figure}

Interestingly, the stars T183 and B2151, which were observed on two consecutive nights to examine potential 
spectrum variability show small changes in radial velocities from one night to the next.
The values $v_{\rm rad}$ in Table~\ref{tab:photometry} are heliocentric radial velocities measured by
cross-correlations. The zero point has been established by using our previous observations of HgMn stars with
FEROS. Dispersion of the subregions indicated an error less than 0.2\,km\,s$^{-1}$, and 
the total absolute error including calibrations and radial velocities of the template
was less than 0.4\,km\,s$^{-1}$. Further the radial velocities were improved by using telluric lines, so 
that relative velocites are better than 0.2-0.3\,km\,s$^{-1}$.
The star T183 shows a radial velocity change of 0.9\,km\,s$^{-1}$ from one night to the next, whereas the 
shift in radial velocity for B2151 amounts to the value 0.4\,km\,s$^{-1}$ which is is of the 
order of the measurement accuracy. We conclude that it is possible that T183 is a spectroscopic binary,
although further spectroscopic observations are highly desirable.

In all stars  complex interstellar structures are found 
close to the position of \ion{Ca}{ii} at 3933.664\,\AA{} and 3968.469\,\AA{} 
and of \ion{Na}{i} at  5889.950\,\AA{} and 5895.924\,\AA{}.
The interstellar structure  affects the blue wings of the \ion{Ca}{ii} lines 
in T181, T191, and T193.
The interstellar lines are blue-shifted with respect to the stellar wavelengths in T181, T191, and T193,
while they are red-shifted in B652, B2151, and B2206. 

\begin{table*}
\caption{Final ATLAS12 abundances. For elements not listed in the table 
the abundances are those of the cluster, i.e.\ [$-$1.91] for NGC\,6397
and [$-$1.61] for NGC\,6752. The elemental abundances are given 
as $\log$(N$_{\rm elem}$/N$_{\rm tot}$).
Similar to the entries in Table 6, errors are given only for lines with measured 
equivalent widths, and no errors are presented for lines
with abundances from comparison of observed and computed profiles
using the synthetic spectrum method. For \ion{Mn}{i} we did not use equivalent
widths because Mn is affected by hyperfine structure. The hyperfine components are
considered in the synthetic spectrum.} 
\label{tab:ATLAS12abundances}
\centering
\begin{tabular}{lcccccccc}
\hline
\noalign{\smallskip}
\multicolumn{1}{c}{Elem}& 
\multicolumn{1}{c}{T183}&
\multicolumn{1}{c}{T191}&
\multicolumn{1}{c}{T193}&
\multicolumn{1}{c}{NGC\,6397}&
\multicolumn{1}{c}{B652}&
\multicolumn{1}{c}{B2151}&
\multicolumn{1}{c}{B2206}&
\multicolumn{1}{c}{NGC\,6752} \\
\hline
\noalign{\smallskip}
       &(11390,3.75) &(11630,3.8) &(11370,3.80)& &(11960,3.95) &(11310,3.85) &(11200,3.80) & \\
\hline
\noalign{\smallskip}
He    & $-$2.07          & $-$1.76          & $-$2.10          & $-$1.11 & $-$2.15          & $-$2.30          & $-$2.30&$-$1.11\\  
O     & $-$5.21          & $-$4.85          & $-$5.51          & $-$5.12 & $-$5.50          & $-$5.50          & $-$5.50&$-$4.82\\
Na    & $-$7.62          & $-$7.62          & $-$7.62          & $-$7.62 & $-$5.90          & $-$7.32          & $-$6.51&$-$7.32\\
Mg    & $-$6.11          & $-$6.38          & $-$6.00          & $-$6.37 & $-$5.68          & $-$5.75          & $-$5.71&$-$6.07\\    
Al    & $-$7.48          & $-$7.48          & $-$7.48          & $-$7.48 & $-$7.18          & $-$7.25          & $-$7.18&$-$7.18\\
Si    & $-$7.49          & $-$6.76          & $-$7.80          & $-$6.40 & $-$7.60          & $-$7.70          & $-$8.00&$-$6.10\\
P     & $-$5.40          & $-$8.50          & $-$5.42$\pm$0.10 & $-$8.50 & $-$4.87$\pm$0.25 & $-$5.53          & $-$5.57$\pm$0.11   &$-$8.20\\
Ca    & $-$5.98           & $-$5.52          & $-$6.22          & $-$7.59 & $-$5.85          & $-$6.34          & $-$6.28&$-$7.29\\
Ti    &$-$6.37$\pm$0.08  & $-$5.38$\pm$0.10 &$-$6.73$\pm$0.09  & $-$8.93 & $-$7.07$\pm$0.11 & $-$6.91$\pm$0.07 & $-$6.91$\pm$0.07   &$-$8.63\\
Cr    &$-$7.06$\pm$0.05  & $-$6.45$\pm$0.03 &$-$6.56$\pm$0.05  & $-$8.28 & $-$6.86$\pm$0.04 & $-$6.90$\pm$0.03 & $-$6.82$\pm$0.05   &$-$7.98\\
Mn    &$-$6.05           & $-$5.50          &$-$5.44           & $-$8.56 & $-$5.65          & $-$6.50          & $-$6.20            &$-$8.26\\
Fe    &$-$4.19$\pm$0.09  & $-$4.69$\pm$0.10 &$-$4.29$\pm$0.10  & $-$6.45 & $-$3.85$\pm$0.11 & $-$4.20$\pm$0.11 & $-$4.22$\pm$0.08   &$-$6.15\\
Co    &$-$9.03           & $-$6.22          &$-$9.03            & $-$9.03 & $-$8.73          & $-$8.73          & $-$8.73  &$-$8.73\\
Ni    &$-$6.14           & $-$6.00          &$-$6.25            & $-$7.70 & $-$6.30          & $-$6.60          & $-$6.60&$-$7.40\\
Y     &$-$8.20           & $-$8.20          &$-$8.19$\pm$0.05   & $-$11.71& $-$7.52$\pm$0.12 &$-$8.50           & $-$8.38$\pm$0.8    &$-$11.41\\
Zr    &$-$11.35          &$-$11.35         &$-$11.35            & $-$11.35 &$-$11.05         &$-$11.05           & $-$9.05        &$-$11.05\\
\hline
\noalign{\smallskip}
\end{tabular}
\end{table*}

\begin{table*}
\caption{Final abundances $\log$(N$_{\rm elem}$/N$_{\rm tot}$) relative to the solar 
abundances from Grevesse \& Sauval (\cite{Grevesse1998}) as derived from ATLAS12 models.} 
\label{tab:rel_abundances}
\centering
\begin{tabular}{rrrrrrr}
\hline
\noalign{\smallskip}
& \multicolumn{3}{c}{NGC\,6397} &
\multicolumn{3}{c}{NGC\,6752} \\
\hline
\noalign{\smallskip}
\multicolumn{1}{c}{Elem.} & 
\multicolumn{1}{c}{T183}& 
\multicolumn{1}{c}{T191} & 
\multicolumn{1}{c}{T193}&
\multicolumn{1}{c}{B652}&
\multicolumn{1}{c}{B2151}&
\multicolumn{1}{c}{B2206} \\
\hline
\noalign{\smallskip}
  &(11390,3.75) &(11630,3.80) &(11370,3.80) &(11960,3.95) &(11310,3.85) &(11200,3.80) \\
\hline
\noalign{\smallskip}
[M/H]    &\multicolumn{3}{c}{[$-$1.91]}  & \multicolumn{3}{c}{[$-$1.61]}\\
\hline
\noalign{\smallskip}
\ion{He} & [$-$0.96]    & [$-$0.65] & [$-$0.99] & [$-$1.04] & [$-$1.19] & [$-$1.19]\\
\ion{O}{i} &$\le$[$-$2.00]& [$-$1.64]&$\le$[$-$2.30] & $\le$[$-$2.29] &$\le$ [$-$2.29] & $\le$[$-$2.29]\\
\ion{Na}{i} &           &           &           & [$-$0.19] &     & $-$[0.80]\\
\ion{Mg}{ii} &[$-$1.65] & [$-$1.92] & [$-$1.54] & [$-$1.22] & [$-$1.29] & [$-$1.25]\\
\ion{Al}{i}  &          &           &           &           &  $\le$[$-$1.68] & \\
\ion{Si}{ii} &[$-$3.00] & [$-$2.27] &$\le$ [$-$3.31] &$\le$ [$-$3.11] &$\le$[$-$3.21 & $\le$[$-$3.51]]\\
\ion{P}{ii} &[$+$1.19]  &           & [$+$1.17] & [$+$1.72] & [$+$1.06] & [$+$1.02]\\
\ion{Ca}{ii} &[$-$0.30] & [$+$0.16] & [$-$0.54] & [$-$0.17] &[$-$0.66] & [$-$0.60] \\
\ion{Ti}{ii} &[$+$0.65] & [$+$1.64] & [$+$0.29] & [$-$0.05] &[$+$0.11] & [$+$0.11]\\
\ion{Cr}{ii} &[$-$0.69] & [$-$0.08] & [$-$0.19] & [$-$0.49] & [$-$0.53] & [$-$0.45]\\
\ion{Mn}{ii} &[$+$0.60] & [$+$1.15] & [$+$1.21] & [$+$1.00] & [$+$0.15] & [$+$0.45]\\
\ion{Fe}{i}-\ion{Fe}{ii} & [$+$0.35] & [$-$0.15] & [$+$0.25] & [$+$0.69] & [$+$0.34] & [$+$0.32]\\
\ion{Co}{ii}&           & [$+$0.90]\\
\ion{Ni}{ii}&[$-$0.35]  & [$-$0.21] & [$-$0.56] & [$-$0.51] & [$-$0.81] & [$-$0.81]\\
\ion{Y}{ii} & [$+$1.60] & [$+$1.60] & [$+$1.61] & [$+$2.28] & [$+$1.30] & [$+$1.42]\\
\ion{Zr}{ii}&           &           &           &           &           & [$+$0.39]\\
%{He} & [$-$0.96]    & [$-$0.65] & [$-$0.99] & [$-$1.04] & [$-$1.19] & [$-$1.19]\\
%\ion{O}{i} &$\le$[$-$2.00]& [$-$1.64]&$\le$[$-$2.30] & $\le$[$-$2.29] &$\le$ [$-$2.29] & $\le$[$-$2.29]\\
%\ion{Na}{i} &           &           &           & [$-$0.19] &     & $-$[0.80]\\
%\ion{Mg}{ii} &[$-$1.65] & [$-$1.92] & [$-$1.54] & [$-$1.22] & [$-$1.29] & [$-$1.25]\\
%\ion{Al}{i}  &          &           &           &           &  $\le$[$-$1.68] & \\
%\ion{Si}{ii} &[$-$3.00] & [$-$2.27] &$\le$ [$-$3.31] &$\le$ [$-$3.11] &$\le$[$-$3.21 & $\le$[$-$3.51]]\\
%\ion{P}{ii} &[$+$1.19]  &           & [$+$1.17] & [$+$1.72] & [$+$1.06] & [$+$1.02]\\
%\ion{Ca}{ii} &[$-$0.20] & [$+$0.16] & [$-$0.54] & [$-$0.17] &[$-$0.66] & [$-$0.60] \\
%\ion{Ti}{ii} &[$+$0.67] & [$+$1.64] & [$+$0.29] & [$-$0.05] &[$+$0.11] & [$+$0.11]\\
%\ion{Cr}{ii} &[$-$0.68] & [$-$0.08] & [$-$0.12] & [$-$0.49] & [$-$0.47] & [$-$0.39]\\
%\ion{Mn}{ii} &[$+$0.60] & [$+$1.30] & [$+$1.21] & [$+$1.00] & [$+$0.15] & [$+$0.45]\\
%\ion{Fe}{i}-{ii} & [$+$0.35] & [$-$0.15] & [$+$0.25] & [$+$0.69] & [$+$0.34] & [$+$0.32]\\
%\ion{Co}{ii}&           & [$+$1.00]\\
%\ion{Ni}{ii}&[$-$0.35]  & [$-$0.21] & [$-$0.46] & [$-$0.51] & [$-$0.81] & [$-$0.81]\\
%\ion{Y}{ii} & [$+$1.60] & [$+$1.60] & [$+$1.61] & [$+$2.28] & [$+$1.30] & [$+$1.42]\\
\hline
\noalign{\smallskip}
\end{tabular}
\end{table*}

%\section{Notes on the observed elements}
%{\it Helium}:
\section{Presence of isotopes and emission lines}
%Comparison of spectral peculiarities with those of HgMn stars and Feige\,86}

\subsection{Isotopes}

\begin{figure}
\begin{center}
\includegraphics[height=0.45\textwidth,angle=90,clip=]{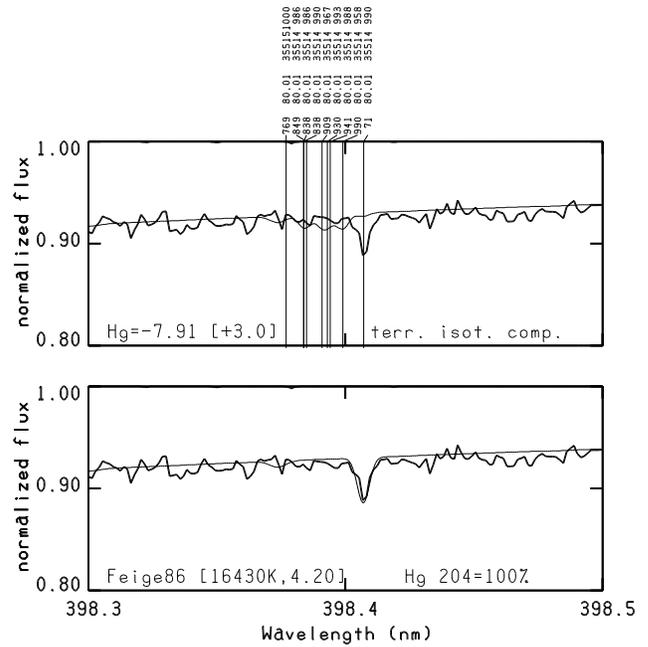}
\end{center}
\caption{
The observation of $^{204}$Hg in the spectrum of Feige\,86.
The upper panel shows the observed UVES spectrum and the synthetic spectrum (thin line)
assuming terrestrial isotopic composition for all isotopes. 
In the bottom panel we show the synthetic spectrum calculated assuming \ion{Hg}{ii} being present 
in the atmosphere entirely in the form of the heaviest stable isotope $^{204}$Hg.
}
\label{fig:feige86}
\end{figure}

The chemical peculiarities in HgMn stars were already studied in numerous papers in the past.
The studies reveal that HgMn stars of similar age, as inferred from their atmospheric parameters, 
can have widely different isotopic anomalies and chemical abundances indicating that the HgMn star 
phenomenon is a result of a complex interplay 
between a number of physical mechanisms (e.g.\ Dolk et al.\ \cite{Dolk2003} ). 
The Pop~II halo B-type star Feige\,86, similar to HgMn stars and BHB shows overabundant 
P, Mn, Fe Ti, Cr and underabundant He.
Apart from the He isotopic anomaly discussed by Bonifacio et al.\ (\cite{Bonifacio95}) 
significant Ca, Pt, and Hg isotopic anomalies 
have recently been discovered in a UVES spectrum of this star (Castelli \& Hubrig, in preparation).
%Pop~II halo B-type star Feige\,86. 
In Fig.~\ref{fig:feige86} we present the observed \ion{Hg}{ii} $\lambda\,3984$ line 
together with the 
synthetic spectrum computed assuming \ion{Hg}{ii} being present in the atmosphere entirely in 
form of the heaviest stable isotope $^{204}$Hg.

Our study of BHB stars revealed no signs of He isotopic anomalies which would produce a
$^{3}$He/$^{4}$He ratio different from the solar one. The largest isotopic
shift of 0.50\,\AA{} separates $^{4}$He  at 6678.152\,\AA{} from
$^{3}$He at 6678.652\,\AA{}. The  $^{3}$He isotope  was not 
observed in any star. In B2151 the line at 6678.152\,\AA{} is so 
weak that an eventual isotope would be merged in the 
noise.
In one BHB star, T191, the line \ion{He}{i} 4471.5\,\AA{} appears blueshifted by $-$0.04\,\AA{}, 
while the line \ion{He}{i} 4713.2\,\AA{} was found redshifted by $+$0.06\,\AA{}.

\begin{figure}
\begin{center}
\includegraphics[width=0.45\textwidth,angle=0,clip=]{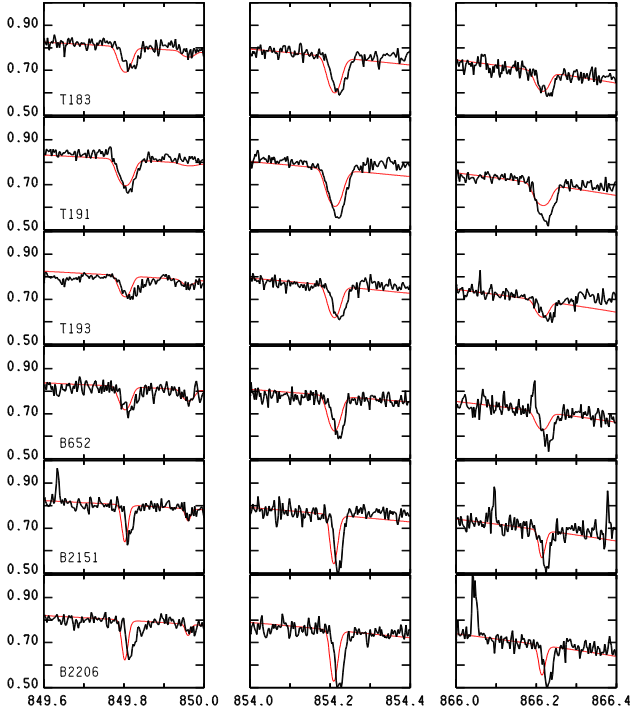}
\end{center}
\caption{
Observed \ion{Ca}{ii} triplet shifts in all studied stars. The computed profiles assuming  
the terrestrial isotopic mixture are presented by thin lines.
}
\label{fig:ca_triplet}
\end{figure}

While in many CP stars isotopic structures of 
Ga and Ba lines as well as  
isotopic anomalous ratios for \ion{He}{i}, \ion{Ca}{ii},
\ion{Pt}{ii}, \ion{Hg}{i}, and \ion{Hg}{ii} lines were observed,
our search for the presence of isotopic structures in the studied stars
has led to recognize the presence of an isotopic anomaly
only for the \ion{Ca}{ii} infrared triplet. 
In all studied BHB stars the \ion{Ca}{ii} triplet lines are redshifted by a different amount with the 
lowest shift of 0.06--0.07\,\AA{} in T191 to 0.13\,\AA{} in T183 and T193. We measure in B2206 a shift of 
0.14\,\AA{} in the \ion{Ca}{ii} infrared triplet line 8662.141, but the Ca profiles in this star are 
rather asymmetrical leading to a lower accuracy of these measurements. 
The accuracy of the wavelength calibration in the region of the Ca triplet
is 0.0018\,\AA{} (rms of the thorium lines). On the other hand, the uncertainty of the effective
wavelength of the Ca triplet lines also depends on the signal-to-noise ratio of
the spectra  and on the shape of the line profile.
Only symmetrical, unblended lines the wavelengths can be measured to an accuracy better 
than 0.01\AA{}.  For asymmetrical lines, differences up to 0.03\,\AA{} may occur even for repeated measurements 
of the same line.  
In Fig.~\ref{fig:ca_triplet} we present the comparison of the observed \ion{Ca}{ii} infrared triplet lines 
in the studied BHB stars with the \ion{Ca}{ii} triplet lines computed assuming the terrestrial isotopic mixture.
In the stars with the lowest \vsini{} values, B2151 and B2206, some kind of structure is noticeable in the 
\ion{Ca}{ii} infrared triplet lines.
% observed in their spectra. 
However, due to the rather low S/N of 
our spectra it is premature to suggest that this structure presents the \ion{Ca}{ii} isotopic 
structure, and the observed features in the line profiles could actually be produced by noise. 
Future studies of these \ion{Ca}{ii} infrared triplet lines in B2151 and B2206 would require
much higher S/N spectroscopic material.
Table~\ref{tab:shifts} lists the amount of the observed shifts. 

\begin{table*}
\caption{$\Delta\lambda$ shift for the \ion{Ca}{ii} lines of the infrared triplet.} 
\label{tab:shifts}
\centering
\begin{tabular}{lcccccc}
\hline
\noalign{\smallskip}
&
\multicolumn{1}{c}{T183}&
\multicolumn{1}{c}{T191}&
\multicolumn{1}{c}{T193}&
\multicolumn{1}{c}{B652}&
\multicolumn{1}{c}{B2151}&
\multicolumn{1}{c}{B2206}\\
\multicolumn{1}{c}{$\lambda$(\AA{})}& 
\multicolumn{6}{c}{$\Delta\lambda$(\AA{})} \\
\hline
\noalign{\smallskip}
8498.023 & 0.13 & 0.07 & 0.13 & 0.10 & 0.11 & 0.11\\
8542.091 & 0.11 & 0.06 & 0.13 & 0.11 & 0.11 & 0.11\\
8662.141 & 0.11 & 0.07 & 0.13 & 0.12 & 0.11 & 0.14\\
\hline
\noalign{\smallskip}
\end{tabular}
\end{table*}

\begin{figure}
\begin{center}
\includegraphics[width=0.45\textwidth,angle=0,clip=]{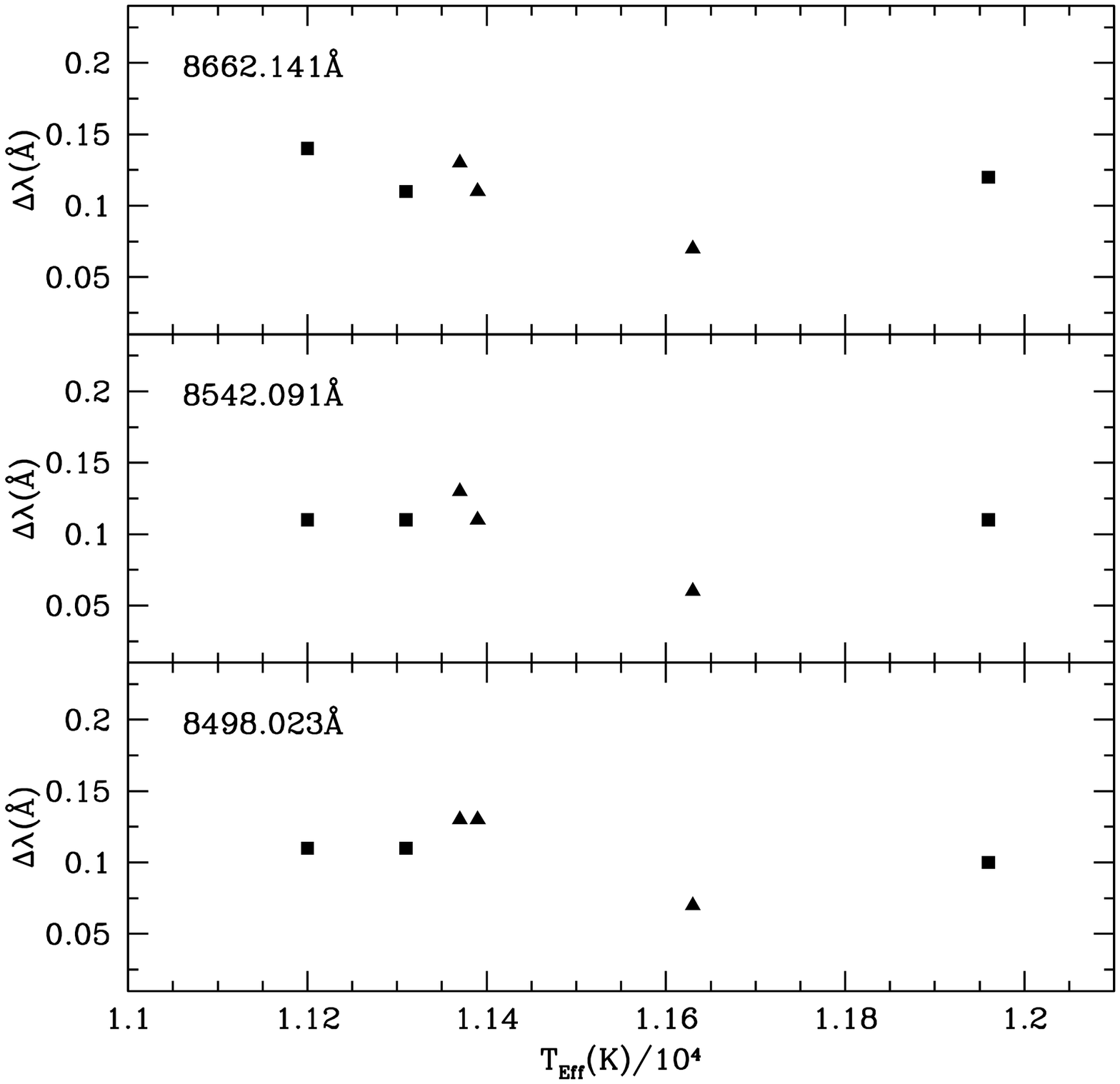}
\includegraphics[width=0.45\textwidth,angle=0,clip=]{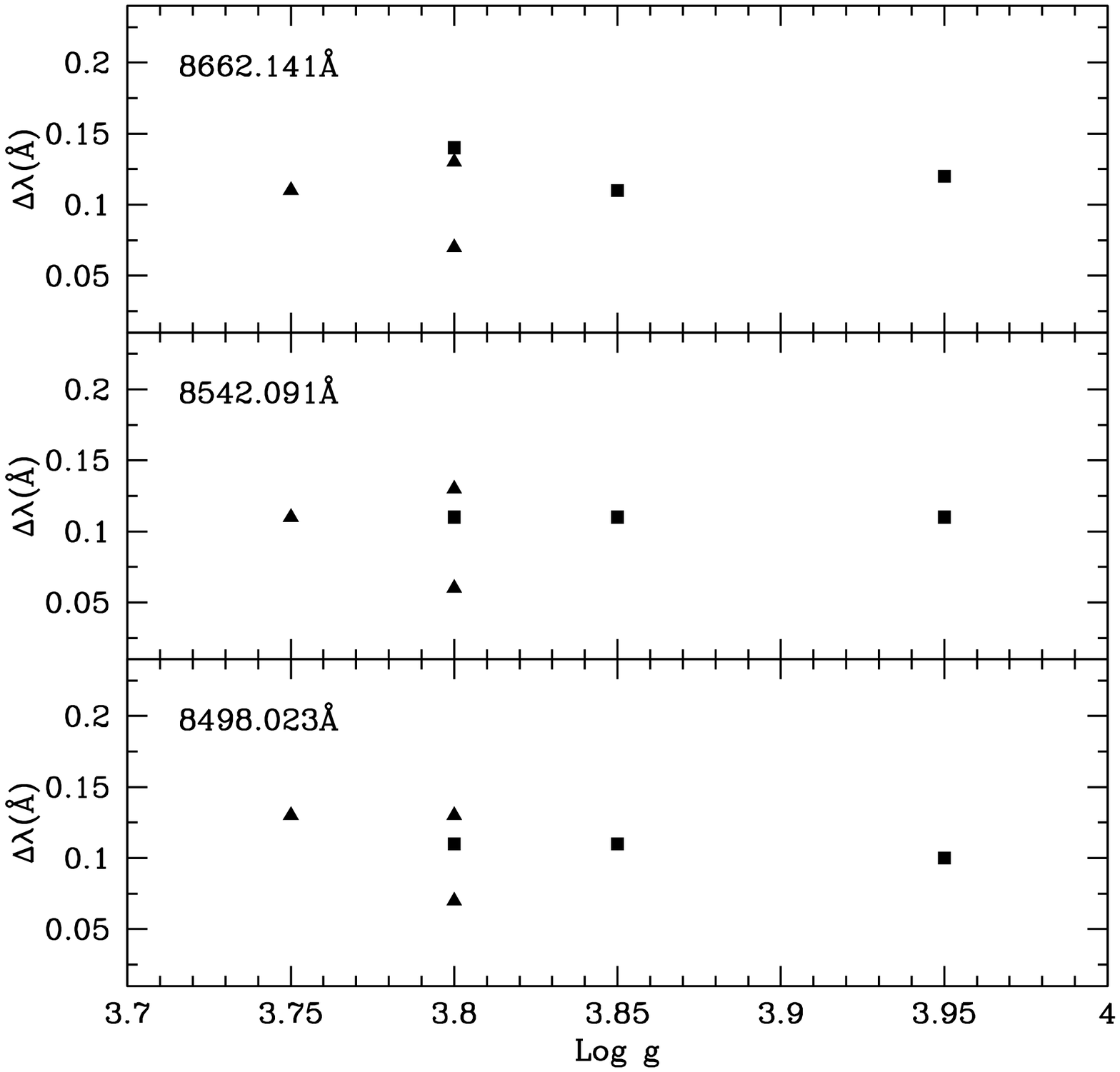}
\end{center}
\caption{
{Ca} isotopic
shifts for each 
studied BHB star for \ion{Ca}{ii} triplet lines ($\lambda$\,8498, $\lambda$\,8542 and $\lambda$\,8662) 
versus  \teff{} and \logg{}. Squares represent BHB stars in NGC\,6397 and triangles represent
BHB stars in NGC\,6752. 
}
\label{fig:Ca_shifts_plots}
\end{figure}

In Fig.~\ref{fig:Ca_shifts_plots} we present for each studied BHB star {Ca} isotopic
shifts for \ion{Ca}{ii} triplet lines ($\lambda$\,8498, $\lambda$\,8542 and $\lambda$\,8662)
versus  \teff{} and \logg{}. We see no indication that Ca isotopic anomalies are related 
to the atmospheric parameters of the studied stars.
Recently, Cowley et al.\ (\cite{Cowley2007}) studied a potential temperature correlation of the Ca isotopic 
anomaly in magnetic Ap stars and HgMn stars. They noticed that there may be a weak correlation with temperature 
for HgMn stars in the sense that hotter stars show larger isotopic shifts. Clearly, more measurements of 
\ion{Ca}{ii} triplet lines in a larger stellar sample are necessary to establish whether a similar trend exists
for BHB stars.

No comparison of isotopic structure with that of HgMn stars can be done for other elements as 
lines of Ga, Ba, Au, Pt, and Hg were not observed in our BHB sample stars.

\subsection{Emissions}

The phenomenon of weak emission lines at optical wavelengths for main-sequence B-type stars and HgMn stars has been noted 
from observations of relatively few stars. Presently, explanations of this phenomenon have been put forward 
in the context of the non-LTE line formation (Sigut \cite{sigut2001}) and possible fluorescence 
mechanisms (Wahlgren \& Hubrig \cite{wahlgren2000}). No obvious correlations has been noted for emission 
in terms of stellar parameters of studied B-type stars. 
In CP stars emissions have been predominantly detected in 
\ion{Mn}{ii}, mult.13, 
\ion{Cr}{ii}, $\lambda\lambda$ 6182.340, 6285.601, 6299.534, 6309.669, 6324.198, 6551.373,
6585.241, 6592.341, 6636.427;
\ion{Ti}{ii}, $\lambda\lambda$ 6001.395, 6029.271, 6040.120, 6042.201, 6106.471,
\ion{Fe}{ii}, $\lambda\lambda$ 6082.882, 6165.893, 6214.948, 6291.830, 6375.792, 8490.079, 8609.506, 8636.587, 8740.349, 8833.963,  
(e.g., Wahlgren \& Hubrig \cite{wahlgren2000}; Wahlgren \& Hubrig \cite{wahlgren2004};
Castelli \& Hubrig \cite{CastelliHubrig2004b}).

In T191 a few emissions are detected in \ion{Ti}{ii} lines in the region 6000-6110\,\AA{} (see e.g. 
Fig.~\ref{fig:emis_ti}).
%while all \ion{Mn}{ii} mult.13 observed lines are much weaker than computed.
It is quite possible that in B652 the 
line \ion{Mn}{ii} $\lambda$ 6125.863 shows a weak emission, but the low signal-to-noise ratio 
does not allow us to consider this line as a detection.
It is also possible that very weak emissions are visible in the spectrum of T183 in lines 
\ion{Mn}{ii} $\lambda\lambda$ 6122.434, 6125.863, 6128.734, 
6130.796, (see e.g. Fig.~\ref{fig:emis_mn}).
For these both stars additional, much higher signal-to-noise observations are required to confirm 
thei presence of   weak emissions in \ion{Mn}{ii} lines. 
\begin{figure}
\begin{center}
\includegraphics[width=0.45\textwidth,angle=0,clip=]{p6039_6045Ti.epsi}
\end{center}
\caption{
Emission line of \ion{Ti}{ii} 6040.120 detected in T191. This line is very weak or absent in other studied BHB stars.
For comparison, we present in the lower panel the same spectral region in T193 where no \ion{Ti}{ii} 6040.120
is visible.}
\label{fig:emis_ti}
\end{figure}

\begin{figure}
\begin{center}
\includegraphics[width=0.45\textwidth,angle=0,clip=]{p6125_6131mn.epsi}
\end{center}
\caption{
A probable presence of \ion{Mn}{ii} 6125.863 emission in T183. This line appears in absorption or is absent 
in other studied BHB stars. For comparison, we present in the lower panel the same spectral region in T193 where
\ion{Mn}{ii} 6125.863 appears in absorption.
}
\label{fig:emis_mn}
\end{figure}

No \ion{Mn}{ii} emission lines were detected in T191 and T193, although the observed lines belonging to mult.13 
are much weaker than computed.
%In T193 there are no emissions. They can be eventually present 
%in \ion{Mn}{ii}(13) whose observed lines are much weaker than computed.
No emission lines were found in other BHB stars. 
%could beIn B652 there are no emissions. Eventually only in \ion{Mn}{ii} 6125.863\,\AA{}.
%There are no emission in B2151 and B2206.

%\section{Interstellar lines}
%\section{The stars}
%\subsection{B652}
 
\section{Discussion}

The main result of the presented observations of two globular clusters is the overabundance 
over the solar values of P, Ti, Mn, Fe, and Y, and the
overabundance over the cluster metallicity of Mg, Ca, and Cr.
This behaviour is the same in all 
six studied stars of both clusters, apart from phosphorus which was not observed in T191.
Furthermore, the Na abundance 
%in B652 and B2206 
is by more than 1.4\,dex larger than the cluster metallicity in B652, and 
by more than 0.8\,dex larger than the cluster metallicity 
in B2206; 
%Na abundance in B652 and B2206 is more than 0.8\,dex larger than the cluster metallicity, 
the Co abundance is 1.0\,dex over the solar abundance for T191, while Zr is overabundant  over the solar 
abundance by 0.4\,dex in B2206.
The Pop~II halo B-type star Feige\,86, similar to BHB stars shows overabundant 
P, Ti, Cr, Mn, and Fe and underabundant He.
A significant Ca and Hg isotopic anomaly has recently been discovered by Castelli \& Hubrig 
(in preparation) in a UVES spectrum of this star.
However, no lines of Hg or other heavy elements were observed in our sample stars.

A vertical abundance inhomogeneous distribution was found in HgMn and in other
subgroups of CP stars. The Ca stratification and isotope separation in 
several chemically peculiar stars was recently investigated by Cowley et
al. (\cite{Cowley2007}, \cite{Cowley2009}) and by Ryabchikova et
al. (\cite{Ryabchikova2008} who showed that in a few stars heavy Ca isotopes 
are most likely concentrated towards the higher atmospheric layers.
%Khalack et al.\ (\cite{Khalack2008}) found signatures of vertical stratification of iron 
%in three BHB stars, B267, B279 and WF2-2541, and of titanium in B267. 
No sign for the presence of an elemental vertical inhomogeneous distribution was found in any spectrum
of the studied stars. In addition, the lines lying in different spectral regions
can be reproduced with same abundance within the limits of the
loggf's accuracy. 

As has been shown by Michaud et al.\ (\cite{Michaud2008}),
the abundance anomalies in BHB stars are believed to be
caused by atomic diffusion, radiative accelerations leading for
instance to the observed Fe overabundances.
The abundance anomalies of Al, Si, S, Ca, Ti, Cr, and Ni, but not of Mg, 
are also reasonably well reproduced by these authors in stellar evolution models which include
all effects of atomic diffusion from the ZAMS to the horizontal branch.
Still, it is not clear yet to what extent the different abundances of the two different 
populations can be explained 
in terms of the different evolutionary history or of different atmospheric phenomena, such 
as diffusion.
% is still an open question.
The link with atomic diffusion is strengthened by the observed slow
rotation, a feature
which is also a characteristic of HgMn stars and is required
to allow the slow diffusion processes to be effective. At least 2/3 
of the known HgMn stars belong to spectroscopic binary systems (Hubrig \& Mathys \cite{HubrigMathys1995}), and 
the slow rotation could in principle be explained with binarity, where 
the components are synchronized with $P_{\rm rot}=P_{\rm orb}$. 
Contrary to HgMn stars, only very few BHB stars are known to belong to a binary system. 
Moni Bidin et al.\ \cite{MoniBidin2006} used the ESO VLT-FORS2 instrument to 
obtain medium resolution (R$\approx$4\,100) spectra of 51 hot HB stars with 8\,000\,K $\le$teff $\le$ 32\,000\,K 
during four consecutive nights, but 
no close binary system has been detected. 
In our study, for both stars, T183 and B2151 observed on two 
consecutive nights small radial velocity changes were found indicating that the fraction of 
binary systems among BHB stars could be larger than estimated in the previous studies. 

In other groups of chemically peculiar stars, magnetic fields are believed to be
responsible for their slow rotation. The presence of global 
magnetic fields contributes to the suppression of turbulence and leads to different
diffusion velocities depending on the field inclination and strength.
To our knowledge, there has been no attempt so
far to detect magnetic fields in BHB stars of globular clusters. Future 
spectropolarimetric studies to establish the
presence or absence of magnetic fields in these stars will provide
important clues about their potential role in slowing them
down.

Furthermore, it is important to obtain high resolution 
high S/N spectra for a larger sample of BHB stars with \teff{} values spanning the range from 
10\,000\,K to much higher \teff{} than in our sample stars, belonging to various globular clusters 
with different metallicity. The gradually increasing number of large telescopes equipped  
with high-resolution spectrographs ensures the availability of such spectroscopic studies. 
They should be used 
to search for a possible correlation between the abundance 
and isotopic anomalies with stellar fundamental parameters to establish the implication of 
diffusion effects on observed atmospheric compositions.

\begin{acknowledgements}
%We are grateful to S.~Moehler for a number of helpful discussions about the choice of suitable targets and 
%providing us with finding charts.
We would like to thank M.~Sch\"oller for his help with the layout of the manuscript and useful suggestions.
\end{acknowledgements}

%MS: check the material after the end of the document

\appendix

\section{Lines used for the abundance analysis}
Table\,A.1 lists the lines that were investigated in the spectra of the six
stars.  The wording ``not obs'' is given for lines not present in the spectra,
while the wordings ``profile'' and ``blend'' are given for lines well observed
in the spectra, but which do not have measurable equivalent widths 
either because the noise affects the profile too much or because 
other components affect the line. These wordings also indicate
lines for which adequate equivalent widths can not be computed 
as in the cases of \ion{Mg}{ii} at 4481\,\AA\ which is 
a blend of transitions belonging to the same multiplet,
of most \ion{Mn}{ii} lines which
are affected by hyperfine structure, of
the \ion{Ca}{ii} infrared triplet which is a blend of isotopic components,
and so on.
For the remaining lines the measured equivelent widths are presented in the table.

\begin{table*}
\caption{Analyzed lines in the stellar spectra and measured equivalent widths in m\AA{}. \ion{He}{i} 
is not included. It is discussed in the text. The \ion{Fe}{i} and \ion{Fe}{ii} lines marked with an asterisk 
were used only to estimate the error on \teff{} due to the particular sample of lines (see Sect. 3.1).} 
\label{tab:analyzed_lines}
\centering
\begin{tabular}{llrrrrrrrrr}
\hline
\hline
\noalign{\smallskip}
\multicolumn{1}{c}{Species} & 
\multicolumn{1}{c}{$\lambda$(\AA{})} & 
\multicolumn{1}{c}{$\log\,gf$} & 
\multicolumn{1}{c}{Ref.} & 
\multicolumn{1}{c}{E$_{i}$ [cm$^{-1}$]}& 
\multicolumn{1}{c}{W(T183)}&
\multicolumn{1}{c}{W(T191)}&
\multicolumn{1}{c}{W(T193)}&
\multicolumn{1}{c}{W(B0652)}&
\multicolumn{1}{c}{W(B2151)}&
\multicolumn{1}{c}{W(B2206)}
  \\
\hline
\noalign{\smallskip}
\ion{C}{ii}& 4267.261 &    0.716 & NIST3 & 145550.700 & not obs & not obs & not obs&not obs& not obs & not obs\\
\ion{N}{i} & 8680.282 &    0.347 & NIST3 &  83364.620 & not obs & not obs & not obs&not obs& not obs & not obs\\
\ion{N}{i} & 8683.403 &    0.087 & NIST3 &  88317.830 & not obs & not obs & not obs&not obs& not obs & not obs\\
\ion{O}{i} & 7771.944 &    0.369 & NIST3 &  73768.200 & not obs & profile & not obs&not obs& not obs & not obs\\
\ion{O}{i} & 7774.166 &    0.223 & NIST3 &  73768.200 & not obs & profile & not obs&not obs& not obs & not obs\\
\ion{O}{i} & 7775.388 &    0.002 & NIST3 &  73768.200 & not obs & profile & not obs&not obs& not obs & not obs\\
\ion{Na}{i}& 5889.950 &    0.108 & NIST3&      0.000 & not obs & not obs & not obs&profile& not obs &profile\\
\ion{Na}{i}& 5895.924 & $-$0.194 & NIST3&      0.000 & not obs & not obs & not obs&profile& not obs &profile?\\
\ion{Mg}{ii}& 4481.126&    0.749 &NIST3 &  71490.190 & profile & profile & profile&profile& profile &profile\\
\ion{Mg}{ii}& 4481.150& $-$0.553 &NIST3 &  71490.190 & profile & profile & profile&profile& profile &profile\\
\ion{Mg}{ii}& 4481.325&    0.594 &NIST3 &  71491.063 & profile & profile & profile&profile& profile &profile\\
\ion{Al}{i} & 3944.006& $-$0.638 &NIST3 &      0.000 & not obs & not obs & not obs&not obs& not obs &not obs\\
\ion{Al}{i} & 3961.520& $-$0.336 &NIST3 &    112.061 & not obs & not obs & not obs&not obs& not obs &not obs\\
\ion{Si}{ii}& 3853.665& $-$1.341 &NIST3 &  55309.350 & not obs & profile & not obs&not obs& not obs &not obs\\
\ion{Si}{ii}& 3856.018& $-$0.406 &NIST3 &  55325.180 & profile & profile & not obs&not obs& not obs &not obs\\
\ion{Si}{ii}& 3862.595& $-$0.757 &NIST3 & 55309.350 & not obs & profile & not obs&not obs& not obs &not obs\\ 
\ion{Si}{ii}& 4130.872& $-$0.783 &NIST3  & 79355.020 & not obs & profile & not obs&not obs& not obs &not obs\\
\ion{Si}{ii}& 4130.894&    0.552 &NIST3  & 79355.020 & not obs & profile & not obs&not obs& not obs &not obs\\
\ion{P}{ii} & 4044.576&    0.481 &K,MRB &107360.250 &profile &not obs &profile & 20.9 & profile& profile\\
\ion{P}{ii} & 4127.559& $-$0.110 &K,KP  &103667.860 &profile &not obs &not obs & 10.6 & profile& profile\\
\ion{P}{ii} & 4420.712& $-$0.329 &NIST3  & 88893.220 &profile &not obs &7.3& 17.7 & not obs & 5.3\\
\ion{P}{ii} & 4475.270&    0.440 &NIST3  &105549.670 &profile &not obs &9.1& 13.1 &not obs& 7.1\\
\ion{P}{ii} & 6024.178&    0.137 &NIST3  & 86743.960 &profile &not obs &profile & 37.9 &profile&profile\\
\ion{P}{ii} & 6034.039& $-$0.220 &NIST3  & 86597.550 &profile &not obs &profile & 27.4 &profile& profile\\ 
\ion{P}{ii} & 6043.084&    0.416 &NIST3  & 87124.600 &profile &not obs &18.8& 42.9 &profile& 12.8\\
\ion{S}{ii} & 4153.068&    0.617 &NIST3  &128233.200 &not obs &not obs &not obs& not obs& not obs & not obs \\
\ion{S}{ii} & 4162.665&    0.777 &NIST3  &128599.160 &not obs &not obs &not obs& not obs& not obs & not obs \\
\ion{Ca}{ii}& 3933.663&    0.135 &NIST3 &     0.000 &profile &profile &profile& profile& profile & profile\\
\ion{Ca}{ii}& 3968.469& $-$0.180 &NIST3 &     0.000 &profile &profile &profile& profile& profile & profile\\
\ion{Ca}{ii}& 8498.023& $-$1.450 &GAL   & 13650.190 &profile &profile &profile& profile& profile & profile\\
\ion{Ca}{ii}& 8542.091& $-$0.500 &GAL   & 13710.880 &profile &profile &profile& profile& profile & profile\\
\ion{Ca}{ii}& 8662.141& $-$0.760 &GAL   & 13650.190 &profile &profile &profile& profile& profile & profile \\
\ion{Sc}{ii}& 4246.822&    0.242 &NIST3 &  2540.950 &not obs &not obs &not obs& not obs& not obs & not obs\\
\ion{Sc}{ii}& 4314.083& $-$0.100 &NIST3 &  4987.790 &not obs &not obs &not obs& not obs& not obs & not obs\\
\ion{Ti}{ii}& 4053.821& $-$1.130 &PTP   & 15265.620 &31.8  & profile &15.3 & profile  & 13.6 & 15.0\\
\ion{Ti}{ii}& 4161.529& $-$2.360 &NIST3 &  8744.250 &profile & 37.8  & profile& not obs &not obs & profile\\
\ion{Ti}{ii}& 4163.644& $-$0.130 &PTP   & 20891.660 &50.1 & 80.6  &38.1 &16.7 & 32.3 & 36.0\\
\ion{Ti}{ii}& 4287.873& $-$1.790 &PTP   &  8710.440 &21.7 & 54.1  & profile & profile & 9.2  & 8.7\\
\ion{Ti}{ii}& 4290.215& $-$0.850 &PTP   &  9395.710 &53.8 & 93.8  &36.6 & 13.6 & 34.5& 34.9\\
\ion{Ti}{ii}& 4294.094& $-$0.930 &PTP   &  9744.250 &55.7 & 83.0  &39.3 & blend & blend & 37.3\\
\ion{Ti}{ii}& 4300.042& $-$0.440 &PTP   &  9518.060 &69.3 &103.8  &58.2 & 31.4 & 49.4& 50.4\\
\ion{Ti}{ii}& 4301.922& $-$1.150 &PTP   &  9363.620 &39.9 & 70.2  &26.5 & 7.8  & 21.5& 21.9\\
\ion{Ti}{ii}& 4312.860& $-$1.100 &PTP   &  9518.060 &51.4 & 78.9  &36.3 & blend  & 26.2& 25.8\\
\ion{Ti}{ii}& 4314.971& $-$1.100 &PTP   &  9363.620 &46.1 & 75.6  &35.7 &blend   & blend & blend\\
\ion{Ti}{ii}& 4367.652& $-$0.860 &PTP   & 20891.660 &28.2 & 59.0  &profile&profile &10.2 & 13.5\\
\ion{Ti}{ii}& 4386.847& $-$0.960 &PTP   & 20951.620 &21.4 & 53.5  &12.4 &not obs & 9.6 & 10.4\\
\ion{Ti}{ii}& 4394.059& $-$1.780 &PTP   &  9850.900 &16.7 & 48.4  & 8.7 &not obs    & 6.4 & 7.1\\
\ion{Ti}{ii}& 4395.031& $-$0.540 &PTP   &  8744.250 &65.1 & 94.0 &53.0 &30.3  &48.1 & 50.9\\
\ion{Ti}{ii}& 4399.765& $-$1.190 &PTP   &  9975.920 &37.0 & 66.4 &24.0 &11.9  &20.4 & 22.1\\
\ion{Ti}{ii}& 4411.072& $-$0.670 &PTP   & 24961.030 &22.2 & 51.6  &15.3 &not obs &10.6 & 12.0\\
\ion{Ti}{ii}& 4417.714& $-$1.190 &PTP   &  9395.710 &37.2 & 70.5  &25.2 &11.6  &22.3 & 22.0\\
\ion{Ti}{ii}& 4418.331& $-$1.970 &PTP   &  9975.920 &12.3 & 42.1  &profile &not obs & 7.1 & 5.8\\
\ion{Ti}{ii}& 4443.801& $-$0.720 &PTP   &  8710.440 &65.6 & 90.6  & 46.5 &24.6 &42.5 & 44.1\\
\ion{Ti}{ii}& 4450.482& $-$1.520 &PTP   &  8744.250 &32.0 & 64.9  &profile &blend  &15.4  &14.1\\
\ion{Ti}{ii}& 4464.448& $-$1.810 &PTP   &  9363.620 &19.0 & 51.8  & 13.0 &not obs &6.8  & 8.3\\
\ion{Ti}{ii}& 4468.492& $-$0.620 &NIST3 &  9118.260 &64.3 & 90.5  & 48.0 & 23.9& 44.0& 45.7\\
\ion{Ti}{ii}& 4488.325& $-$0.510 &PTP   & 25192.710 &28.3 & 59.1  &18.8 &profile &14.9 &16.0\\
\ion{Ti}{ii}& 4501.270& $-$0.770 &PTP   &  8997.710 &72.4 &94.9   &45.8 &26.4 &44.7 & 47.5\\
\ion{Ti}{ii}& 4563.757& $-$0.690 &PTP   &  9050.900 &51.9 & 87.7  &40.0 & 20.2 & 30.8 &37.3\\
\ion{Ti}{ii}& 4571.971& $-$0.320 &PTP   & 12676.970 &71.7 &101.5  &57.3 &33.4  & 49.5 &55.6\\
\ion{Ti}{ii}& 4805.085& $-$1.120 &NIST3 & 16625.110 &33.6 & 64.9  &20.8 &profile  &16.9  &15.5\\
\ion{Ti}{ii}& 4911.195& $-$0.610 &PTP   & 25192.790 &25.0 &59.9   &17.7 &profile &12.1  &12.9\\
\hline
\noalign{\smallskip}
\end{tabular}
\end{table*}
 
\begin{table*}
\addtocounter{table}{-1}
\caption{Cont.} 
\centering
\begin{tabular}{llrrrrrrrrr}
\hline
\noalign{\smallskip}
\multicolumn{1}{c}{Species} & 
\multicolumn{1}{c}{$\lambda$(\AA)} & 
\multicolumn{1}{c}{$\log\,gf$} & 
\multicolumn{1}{c}{Ref.$^{a}$} & 
\multicolumn{1}{c}{E$_{i}$(cm$^{-1}$}& 
\multicolumn{1}{c}{W(T183)}&
\multicolumn{1}{c}{W(T191)}&
\multicolumn{1}{c}{W(T193)}&
\multicolumn{1}{c}{W(B0652)}&
\multicolumn{1}{c}{W(B2151)}&
\multicolumn{1}{c}{W(B2206)}
  \\
\hline
\noalign{\smallskip}
\ion{Cr}{ii} & 4558.650& $-$0.410  & SL  & 32854.310 & 27.3 & 48.5 &47.0 & 30.6 & 32.4 & 36.2\\
\ion{Cr}{ii} & 4588.199& $-$0.640  & NIST3 & 32836.680 & 22.0 & 37.8 &40.2 &blend   & 26.9 & 31.8\\
\ion{Cr}{ii} & 4618.803& $-$0.860  & SL  & 32854.950 & 12.5 & 32.2 &27.3 &13.7  & 18.5 &22.1\\
\ion{Cr}{ii} & 4634.070& $-$0.990  & SL  & 32844.760 & profile & 26.7 &23.7 &profile&13.7  &16.4\\
\ion{Mn}{ii}$^{1}$ & 4206.367& $-$1.590  & K03Mn & 43528.640 &profile&profile&profile&profile& not obs &profile\\
\ion{Mn}{ii} & 4242.333 &$-$1.288  & K03Mn & 49820.160&profile&profile&blend&blend& blend & blend\\
\ion{Mn}{ii} & 4252.963 &$-$1.172  & K03Mn & 49889.860&profile&profile&profile&profile&profile&profile\\
\ion{Mn}{ii} & 4253.025 &$-$2.312  & K03Mn & 43395.380&profile&profile&profile&profile&not obs&profile\\
\ion{Mn}{ii} & 4253.112  &$-$2.079  & K03Mn & 49889.860&profile&profile&profile&profile&not obs&profile\\
\ion{Mn}{ii}$^{1}$ & 4282.490&$-$1.698 & K03Mn & 44521.520&profile&profile&profile&blend&blend&profile\\
\ion{Fe}{i}$^{*}$ & 3815.840 & 0.232   &FW06 & 11976.238 &40.8 & 22.8 & 33.2 &45.2  & 46.2  & 38.7 \\
\ion{Fe}{i}$^{*}$ & 3820.425 & 0.119   &FW06 &  6928.268 &50.4 & 30.3 & 45.2 &46.3  & 55.8  & 49.4 \\
\ion{Fe}{i}$^{*}$ & 3856.371 &$-$1.286 &FW06 &   415.933 &13.0 & profile& profile&profile & 15.3  & 14.8 \\
\ion{Fe}{i}$^{*}$ & 3859.911 &$-$0.710 &FW06 &     0.000 &27.8 & 21.2 & 27.8 & 31.6 & 36.8  & 33.9 \\
\ion{Fe}{i}  & 4005.242       &$-$0.610 &FW06 & 12560.933 &profile &profile & 13.3 & 18.9 & 15.8 & 7.6\\
\ion{Fe}{i}  & 4045.812       &   0.280 &FW06 & 11976.238 &41.9 &20.1 & 38.1 & 42.9 & 44.1 & 45.1\\
\ion{Fe}{i}  & 4071.738       &$-$0.022 &FW06 & 12968.553 &28.5 &profile& 23.5 & 27.8 & 30.1 & 31.9\\
\ion{Fe}{i}  & 4202.029       &$-$0.708 &FW06 & 11976.238 &10.5 &profile&  9.3 & 16.0 & 12.3 & 14.4\\
\ion{Fe}{i}  & 4219.360       &   0.000 &FW06 & 28819.952 & profile&not obs & profile & not obs &  7.1 &  7.0\\
\ion{Fe}{i}  & 4235.936       &$-$0.341 &FW06 & 43163.323 & profile&not obs & profile &profile  & 11.1 & 10.7\\
\ion{Fe}{i}  & 4271.760       &$-$0.164 &FW06 & 11976.238 &29.4 &profile   & 26.9 & 31.9 & 30.5 & 32.2\\
\ion{Fe}{i}  & 4383.545       &   0.200 &FW06 & 11976.238 &43.3 &21.7 & 37.5 & 38.8 & 45.7 & 41.5\\
\ion{Fe}{i}  & 4404.750       &$-$0.142 &FW06 & 12560.933 &26.2 &profile& 27.4 & 29.3 & 29.2 & 32.3\\
\ion{Fe}{i}  & 4415.122       &$-$0.615 &FW06 & 12968.553 &17.6 &profile& 14.2 & 18.1 & 16.6 & 17.2\\
\ion{Fe}{ii} & 4122.668       &$-$3.300 &FW06 & 20830.582 &39.4 &20.1 & 36.6 & 47.2 & 39.5 & 37.8\\
\ion{Fe}{ii} & 4128.748       &$-$3.578 &FW06 & 20830.582 &24.5 &13.7 & 23.2 & 33.5 & 25.9 & 27.9\\
\ion{Fe}{ii} & 4178.862       &$-$2.443 &FW06 & 20830.582 &59.6 &46.8 & 56.1 & 70.4 & 61.1 & 66.5\\
\ion{Fe}{ii} & 4258.154       &$-$3.480 &FW06 & 21812.055 &31.0 &14.3 & 27.2: & 38.8 & 31.7 & 29.2\\
\ion{Fe}{ii} & 4273.326       &$-$3.300 &FW06 & 21812.055 &32.8 &18.5 & 30.9 & 43.6 & 33.9 & 34.1\\
\ion{Fe}{ii} & 4296.572       &$-$2.930 &FW06 & 21812.055 &50.2 &28.3 & 42.9 & 58.1 & 46.9 & 47.3\\
\ion{Fe}{ii} & 4303.176       &$-$2.610 &FW06 & 21812.055 &65.5 &49.0 & 59.4 & 68.6 & 61.4 & 61.9\\
\ion{Fe}{ii} & 4369.411       &$-$3.580 &FW06 & 22409.852 &27.9 &13.0 & 22.7 & 33.8 & 25.3 & 26.6\\
\ion{Fe}{ii} & 4385.387       &$-$2.580 &FW06 & 22409.852 &62.7 &37.5 & 57.0 & 63.6 & 57.7 & 57.3\\
\ion{Fe}{ii} & 4413.601       &$-$4.190 &FW06 & 21581.638 &10.0 &not obs & 12.7 & 19.1 & 14.8 & 11.7\\
\ion{Fe}{ii} & 4416.830       &$-$2.600 &FW06 & 22409.852 &57.1 &38.9 & 55.3 & 65.6 & 63.0 & 58.9\\
\ion{Fe}{ii} & 4491.405       &$-$2.640 &FW06 & 23031.300 &53.7 &34.5 & 50.7 & 64.0 & 52.5 & 52.3\\
\ion{Fe}{ii} & 4508.288       &$-$2.350 &FW06 & 23031.300 &69.1 &48.9 & 65.2 & 77.4 & 68.6 & 68.1\\
\ion{Fe}{ii} & 4515.339       &$-$2.360 &FW06 & 22939.358 &64.8 &43.8 & 58.1 & 68.8 & 62.7 & 65.1\\
\ion{Fe}{ii} & 4520.224       &$-$2.620 &FW06 & 22637.205 &55.2 &40.5 & 52.3 & 63.5 & 57.6 & 56.9\\
\ion{Fe}{ii} & 4522.634       &$-$1.990 &FW06 & 22939.358 &69.5 &54.6 & 69.4 & 81.3 & 70.1 & 74.4\\
\ion{Fe}{ii} & 4541.524       &$-$2.970 &FW06 & 23031.300 &47.0 &27.3 & 41.5 & 54.0 & 48.3 & 45.7\\
\ion{Fe}{ii} & 4549.192       &$-$1.705 &K07F & 47674.721 &39.8 &21.7 & 36.3 & profile  & 37.1 & 37.3\\
\ion{Fe}{ii} & 4549.474       &$-$1.730 &FW06 & 22810.357 &blend  & blend & blend  & blend  & 82.3:& 81.6:\\
\ion{Fe}{ii} & 4555.893       &$-$2.250 &FW06 & 22810.357 &69.2 &49.4 & 61.0 & 69.2 & 64.6 & 64.5\\
\ion{Fe}{ii} & 4576.340       &$-$2.920 &FW06 & 22939.358 &47.5 &28.7 & 41.0 & 56.3 & 44.4 & 46.4\\
\ion{Fe}{ii} & 4582.835       &$-$3.062 &FW06 & 22939.358 &41.7 &21.6 & 37.7 & 45.6 & 39.2 & 37.4\\
\ion{Fe}{ii} & 4583.837       &$-$1.740 &FW06 & 22637.205 &88.9 &63.8 & 86.2 & 102.3& 82.1 & 82.1\\
\ion{Fe}{ii}$^{*}$ & 6147.741 &$-$2.702 &K07F & 31364.440 &32.4 &20.0 & 26.690 & 49.4 & 39.2 & 33.6\\
\ion{Fe}{ii}$^{*}$ & 6149.258 &$-$2.703 &K07F & 31368.450 &35.0 &bad spect.&29.8& 45.5& 34.1 & 36.7\\
\ion{Fe}{ii}$^{*}$ & 6383.722 &$-$2.235 &K07F & 44784.761 &17.6 &not obs &14.8 & 25.9 &19.1  & 19.8\\
\ion{Fe}{ii}$^{*}$ & 6416.928 &$-$2.718 &K07F & 31387.948 &38.0 &18.1 &28.4  & 43.7 &30.8  & 28.1\\
\ion{Fe}{ii}$^{*}$ & 6456.383 &$-$2.057 &K07F & 31483.176 &60.7 &40.8 &53.8  &73.7 & 56.4  & 56.6\\ 
\ion{Co}{ii} & 4160.657       &$-$1.751 &K06  & 27484.371 &not obs&profile&not obs & not obs & not obs & not obs\\
\ion{Ni}{ii} & 4015.474       &$-$2.410 &K03Ni  & 32523.540 &profile&profile&profile& profile& not obs &profile\\
\ion{Ni}{ii} & 4067.031       &$-$1.834 &K03Ni  & 32499.530 &profile&profile&profile& profile&profile& profile\\
\ion{Ga}{ii} & 4255.722       &   0.634 &RS94 &113842.190 &not obs&not obs&not obs&not obs & not obs& not obs \\
\ion{Ga}{ii} & 6334.069       &   1.000 &RS94 &102944.550 &not obs&not obs&not obs&not obs & not obs&not obs\\
\ion{Sr}{ii} & 4077.709       &   0.151 &NIST3 &     0.000 &not obs&not obs&not obs&not obs & not obs&not obs\\
\ion{Y}{ii}  & 3950.349       &$-$0.490 &K,HL   & 840.213  &profile & profile  &12.6  &25.1 & profile &profile   \\
\ion{Y}{ii}  & 4883.682       &   0.070 &K,HL   &8743.316  &profile & profile  &17.2  &21.0 & profile & 11.9\\
\ion{Y}{ii}  & 4900.120       &$-$0.090 &K,HL   &8328.041  &profile & profile &12.3  &24.6 &profile   & 12.3\\
\hline
\noalign{\smallskip}
\end{tabular}
\end{table*}

\begin{table*}
\addtocounter{table}{-1}
\caption{Cont.} 
\begin{flushleft}
\centering
\begin{tabular}{rrrrrrrrrrr}
\hline
\noalign{\smallskip}
\multicolumn{1}{c}{Species} & 
\multicolumn{1}{c}{$\lambda$(\AA)} & 
\multicolumn{1}{c}{$\log\,gf$} & 
\multicolumn{1}{c}{Ref.$^{a}$} & 
\multicolumn{1}{c}{E$_{i}$(cm$^{-1}$}& 
\multicolumn{1}{c}{W(T183)}&
\multicolumn{1}{c}{W(T191)}&
\multicolumn{1}{c}{W(T193)}&
\multicolumn{1}{c}{W(B0652)}&
\multicolumn{1}{c}{W(B2151)}&
\multicolumn{1}{c}{W(B2206)}
  \\
\hline
\noalign{\smallskip}
\ion{Zr}{ii}& 4149.198 & $-$0.040 & LNAJ   &  6467.610 & not obs& not obs&not obs&not obs& not obs&profile\\
\ion{Xe}{ii}& 4603.005 &    0.017 & NIST3  & 95064.380 & not obs& not obs&not obs&not obs& not obs&not obs \\
\ion{Xe}{ii}& 4844.330 &    0.491 & NIST3  & 93068.440 & not obs& not obs&not obs&not obs& not obs&not obs \\
\ion{Ba}{ii}& 4554.029 & $-$0.161 & NIST3  &     0.000 & not obs& not obs&not obs&not obs& not obs&not obs\\
\ion{Ba}{ii}& 4934.076 & $-$0.458 & NIST3  &     0.000 & not obs& not obs&not obs&not obs& not obs &not obs\\
\ion{Pt}{ii}& 4514.124 & $-$1.480 & DSJ    & 29261.970 & not obs& not obs&not obs&not obs& not obs &not obs\\
\ion{Au}{ii}& 4016.067 & $-$1.880 & RW     & 48510.890 & not obs& not obs&not obs&not obs& not obs &not obs\\
\ion{Au}{ii}& 4052.790 & $-$1.690 & RW     & 48510.890 & not obs& not obs&not obs&not obs& not obs &not obs\\
\ion{Hg}{ii}& 3983.890 & $-$1.510 & NIST3  & 35514.000 & not obs& not obs&not obs&not obs& not obs&not obs\\
\hline
\noalign{\smallskip}
\end{tabular}
\end{flushleft}
$^{1}$ The hyperfine structure was considered in the line profile computations.\\
$^{a}$ (NIST3) NIST Atomic Spectra Database, version 3 at http://physics.nist.gov;\\
(DSJ) Dworetsky et al. (1984);\\
(FW06) Fuhr \& Wiese (2006);\\
(GAL) Gallager (1967); \\
(LNAJ) Ljung et al. (2006);\\
(PTP) Pickerin et al. (2002);\\ 
(RS94) Ryabchikova \& Smirnov (1994);\\ 
(RW) Rosberg \& Wyart (1997);\\
(SL) Sigut \& Landstreet (1990);\\
K03Mn: http://kurucz.harvard.edu/atoms/2501/gf2501.pos;\\
K03Ni: http://kurucz.harvard.edu/atoms/2801/gf2801.pos;\\
K06: http://kurucz.harvard.edu/atoms/2701/gf2701.pos;\\
K07F:http://kurucz.harvard.edu/atoms/2601/gf2601.pos. For these lines the
Kurucz $\log\,gf$'s give a better agreement with the observations than the FW06 $\log\,gf's$;\\
``K'' before another $\log\,gf$ source means that the $\log\,gf$ is from Kurucz files
available at http://kurucz.harvard.edu/linelists/gf100/; 
(HL) Hannaford et al. (1982);(KP) Kurucz\& Peytremann (1975); (MRB) Miller et al. (1971).\\

\end{table*}
\end{document}